\pdfoutput=1
\documentclass[a4paper,cits]{JINST}
\usepackage{lineno}
\usepackage{microtype}
\usepackage{booktabs}
\usepackage{amsmath}
\usepackage[hyphens]{url}

\newcommand{\Edep}{\ensuremath{E_{dep}}}
 
\newcommand{\Ni}{\ensuremath{N_{i}}} 
\newcommand{\Nex}{\ensuremath{N_{ex}}} 
\newcommand{\Wi}{\ensuremath{W_{i}}} 
\newcommand{\Wex}{\ensuremath{W_{ex}}} 
\newcommand{\W}{\ensuremath{W}} 

\newcommand{\Ne}{\ensuremath{N_{e}}} 
\newcommand{\Nph}{\ensuremath{N_{ph}}} 

\newcommand{\ELGain}{\ensuremath{\eta_{EL}}} 

\newcommand{\None}{\ensuremath{N_{1}}} 
\newcommand{\Ntwo}{\ensuremath{N_{2}}}

\newcommand{\Eff}{\ensuremath{\varepsilon}}  

\newcommand{\Rn}[1]{\ensuremath{{}^{#1}}Rn}
\newcommand{\Po}[1]{\ensuremath{{}^{#1}}Po}
\newcommand{\Ra}[1]{\ensuremath{{}^{#1}}Ra}
\newcommand{\Tl}[1]{\ensuremath{{}^{#1}}Tl}
\newcommand{\Bi}[1]{\ensuremath{{}^{#1}}Bi}

\newcommand{\bbznu}{\ensuremath{\beta\beta 0\nu}}

\title{An improved measurement of electron-ion recombination in high-pressure xenon gas}

\author{
\mbox{The NEXT Collaboration}

L.~Serra,$^{a}$\thanks{Corresponding author} ~
M.~Sorel,$^{a}$
V.~\'Alvarez,$^{a}$
F.I.G.~Borges,$^{b}$
M.~Camargo,$^{c}$
S.~C\'arcel,$^{a}$
S.~Cebri\'an,$^{d}$
A.~Cervera,$^{a}$
C.A.N.~Conde,$^{b}$
T.~Dafni,$^{d}$
J.~D\'iaz,$^{a}$
R.~Esteve,$^{e}$
L.M.P.~Fernandes,$^{f}$
P.~Ferrario,$^{a}$
A.L.~Ferreira,$^{g}$
E.D.C.~Freitas,$^{f}$
V.M.~Gehman,$^{h}$
A.~Goldschmidt,$^{h}$
J.J.~G\'omez-Cadenas,$^{a}$\thanks{Spokesperson} ~
D.~Gonz\'alez-D\'iaz,$^{d}$
R.M.~Guti\'errez,$^{c}$
J.~Hauptman,$^{i}$
J.A.~Hernando Morata,$^{j}$
D.C.~Herrera,$^{d}$
I.G.~Irastorza,$^{d}$
L.~Labarga,$^{k}$
A.~Laing,$^{a}$
I.~Liubarsky,$^{a}$
N.~Lopez-March,$^{a}$
D.~Lorca,$^{a}$
M.~Losada,$^{c}$
G.~Luz\'on,$^{d}$
A.~Mar\'i,$^{e}$
J.~Mart\'in-Albo,$^{a}$
G.~Mart\'inez-Lema,$^{j}$
A.~Mart\'inez,$^{a}$
T.~Miller,$^{h}$
F.~Monrabal,$^{a}$
M.~Monserrate,$^{a}$
C.M.B.~Monteiro,$^{f}$
F.J.~Mora,$^{e}$
L.M. Moutinho,$^{g}$
J.~Mu\~noz~Vidal,$^{a}$
M.~Nebot-Guinot,$^{a}$
D.~Nygren,$^{l}$
C.A.B.~Oliveira,$^{h}$
J.~P\'erez,$^{k}$
J.L.~P\'erez~Aparicio,$^{m}$
M.~Querol,$^{a}$
J.~Renner,$^{a}$
L.~Ripoll,$^{n}$
A.~Rodr\'iguez,$^{d}$
J.~Rodr\'iguez,$^{a}$
F.P.~Santos,$^{b}$
J.M.F.~dos~Santos,$^{f}$
D.~Shuman,$^{h}$
A.~Sim\'on,$^{a}$
C.~Sofka,$^{o}$
J.F.~Toledo,$^{e}$
J.~Torrent,$^{n}$
Z.~Tsamalaidze,$^{p}$
J.F.C.A.~Veloso,$^{g}$
J.A.~Villar,$^{d}$
R.~Webb,$^{o}$
J.T.~White,$^{o}$\thanks{Deceased} ~
N.~Yahlali$^{a}$\\
\llap{$^{a}$}
Instituto de F\'isica Corpuscular (IFIC), CSIC \& Universitat de Val\`encia\\
Calle Catedr\'atico Jos\'e Beltr\'an, 2, 46980 Paterna, Valencia, Spain\\
\llap{$^{b}$}
LIP and Departamento de F\'isica, Universidade de Coimbra\\
3004-516 Coimbra, Portugal\\
\llap{$^{c}$}
Centro de Investigaciones, Universidad Antonio Nari\~no\\ 
Carretera 3 este No.\ 47A-15, Bogot\'a, Colombia\\
\llap{$^{d}$}
Lab.\ de F\'isica Nuclear y Astropart\'iculas, Universidad de Zaragoza\\ 
Calle Pedro Cerbuna, 12, 50009 Zaragoza, Spain\\
\llap{$^{e}$}
Instituto de Instrumentaci\'on para Imagen Molecular (I3M), Universitat Polit\`ecnica de Val\`encia\\ 
Camino de Vera, s/n, Edificio 8B, 46022 Valencia, Spain\\
\llap{$^{f}$}
LIBPhys, Physics Department, University of Coimbra\\
Rua Larga, 3004-516 Coimbra, Portugal\\
\llap{$^{g}$}Institute of Nanostructures, Nanomodelling and Nanofabrication (i3N), Universidade de Aveiro\\
Campus de Santiago, 3810-193 Aveiro, Portugal\\
\llap{$^h$}
Lawrence Berkeley National Laboratory (LBNL)\\
1 Cyclotron Road, Berkeley, California 94720, USA\\
\llap{$^{i}$}
Department of Physics and Astronomy, Iowa State University\\
12 Physics Hall, Ames, Iowa 50011-3160, USA\\
\llap{$^{j}$}
Instituto Gallego de F\'isica de Altas Energ\'ias (IGFAE), Univ.\ de Santiago de Compostela\\
Campus sur, R\'ua Xos\'e Mar\'ia Su\'arez N\'u\~nez, s/n, 15782 Santiago de Compostela, Spain\\
\llap{$k$}
Departamento de F\'isica Te\'orica, Universidad Aut\'onoma de Madrid\\
Ciudad Universitaria de Cantoblanco, 28049 Madrid, Spain\\
\llap{$l$}
University of Texas, Arlington \\
Texas 76019, USA\\
\llap{$^{m}$}
Dpto.\ de Mec\'anica de Medios Continuos y Teor\'ia de Estructuras, Univ.\ Polit\`ecnica de Val\`encia\\
Camino de Vera, s/n, 46071 Valencia, Spain\\
\llap{$^{n}$}
Escola Polit\`ecnica Superior, Universitat de Girona\\
Av.~Montilivi, s/n, 17071 Girona, Spain\\
\llap{$^{o}$}
Department of Physics and Astronomy, Texas A\&M University\\
College Station, Texas 77843-4242, USA\\
\llap{$^{p}$}
Joint Institute for Nuclear Research (JINR)\\
Joliot-Curie 6, 141980 Dubna, Russia\\
E-mail: \email{luis.serra@ific.uv.es}
}

\abstract{
We report on results obtained with the NEXT-DEMO prototype of the NEXT-100 high-pressure xenon gas time projection chamber (TPC), filled with pure xenon gas at 10 bar pressure and exposed to an alpha decay calibration source. Compared to our previous measurements with alpha particles, an upgraded detector and improved analysis techniques have been used. We measure event-by-event correlated fluctuations between ionization and scintillation due to electron-ion recombination in the gas, with correlation coefficients between -0.80 and -0.56 depending on the drift field conditions. By combining the two signals, we obtain a 2.8\% FWHM energy resolution for 5.49~MeV alpha particles and a measurement of the optical gain of the electroluminescent TPC. The improved energy resolution also allows us to measure the specific activity of the radon in the gas due to natural impurities. Finally, we measure the average ratio of excited to ionized atoms produced in the xenon gas by alpha particles to be $0.561\pm 0.045$, translating into an average energy to produce a primary scintillation photon of $W_{\rm ex}=(39.2\pm 3.2)$~eV.     
}


\begin{document}


\tableofcontents

\section{Introduction} \label{sec:Introduction}

Xenon time projection chambers (Xe TPCs) are widely used in particle and astroparticle physics. Applications include neutrinoless double beta decay \cite{Auger:2012gs,Alvarez:2012sma}, dark matter direct detection \cite{Aprile:2011dd,Akerib:2012ys} and gamma-ray astrophysics \cite{Aprile:2008ft}. These detectors typically detect both the ionization and the prompt scintillation light produced in xenon after the passage of ionizing particles. The charge plus light dual readout provides not only the full imaging of the energy deposition within the TPC, but also a mean for electron versus nuclear recoil discrimination and for improved detector energy resolution. This is possible thanks to the interplay between the charge and light signals caused by electron-ion recombination in xenon, which is the main focus of this work.

Electron-ion recombination refers to the microscopic process by which the produced ionization electrons recombine with positively-charged xenon ions into neutral xenon atoms, before being drifted by an external electric field toward charge readout planes. Two types of recombination processes are considered. Initial (or geminate) recombination \cite{Onsager:1938zz} refers to electrons recombining with their parent ions. If electrons recombine with a different ion in the surrounding xenon volume, one speaks about volume (or columnar) recombination \cite{jaffe}. In each recombination process, one additional scintillation photon is produced, at the expense of one ionization electron being lost \cite{Aprile:2009dv}. The amount of recombination depends on xenon density, external electric field applied, and particle energy loss per unit path length. An anti-correlation between the average charge and light signals per deposited energy is therefore expected as a function of any of these three quantities. The realization that an average increase in scintillation light as the electric field is lowered can compensate a corresponding average decrease in charge was first observed in liquid xenon in 1978 \cite{Kubota:1978oha}. This same effect has subsequently been observed also in high-pressure xenon gas, see \cite{pushkin, bolotnikov1999, kobayashi, Saito:2003dz}. For highly-ionizing tracks such as nuclear recoils or alpha particles, more recombination occurs compared to electron tracks, decreasing the amount of charge while proportionally increasing the amount of light, see for example \cite{Aprile:2012nq,Akerib:2013tjd}. Because of the stochastic nature of the recombination process (due in part to fluctuations in ionization density), the anti-correlation between charge and light signals is also present on a track-by-track basis. This is used in liquid xenon experiments to improve the energy resolution for electrons, see for example \cite{Albert:2013gpz,Aprile:2011dd}. The same strategy can be used in high-pressure xenon gas detectors. Because of the lower recombination rate at these lower densities, the anti-correlation has only been observed in gas for highly ionizing alpha particles \cite{Alvarez:2012hu}, where this effect is enhanced compared to electrons. This work is based on the same NEXT-DEMO high-pressure xenon gas TPC and on the same alpha calibration source as in \cite{Alvarez:2012hu}. However, the results presented here represent a significant improvement over those in \cite{Alvarez:2012hu}, stemming from an upgraded detector and a more detailed data analysis.

\begin{table}[t!b!]
\caption{Values of the average energy \Wex\ required to produce one excited atom, for high-pressure xenon gas. The experimental conditions of the measurements are also summarized.}\label{tab:Wex}
\begin{center}
\begin{tabular}{cccccc}
\hline
Gas pressure    & Drift field         & Source type & Source energy       & \Wex\          & Reference \\
(bar)           & (kV/cm)             &             & (keV)               & (eV)           &           \\ \hline
1               & 0.35                & X-ray       & 5.9                 & $111\pm 16$    & \cite{docarmo} \\
1--3            & 0.15--0.6           & X-ray       & 5.9                 & $72\pm 6$      & \cite{Fernandes:2010gg} \\
15              & 1.5                 & gamma       & 60                  & $76\pm 12$     & \cite{Parsons:1990hv} \\
14              & 0.37                & gamma       & 662                 & $61.4\pm 18.0$ & \cite{Renner:2014mha} \\
1--10           & 0.3--1.3            & alpha       & 5,490               & $34.5\pm 1.4$  & \cite{Saito:2003dz} \\ 
5               & 0.5                 & alpha       & 5,490               & $34.1\pm 2.4$  & \cite{mimura} \\ \hline
\end{tabular}
\end{center}
\end{table}

The NEXT-DEMO detector is ideally suited also to improve our understanding of the partitioning of energy loss between ionization and excitation in xenon gas. The ratio of excitation to ionization quanta produced in xenon by ionizing radiation, \Nex/\Ni, is customarily used to describe this energy partitioning. This ratio is completely specified if both the absolute ionization and excitation yields are known. If we define \Wi\ (\Wex) as the average energy required to produce one electron-ion pair (one excited atom), it immediately follows that \Nex/\Ni\ is equal to \Wi/\Wex. The absolute ionization yield \Wi\ has been measured accurately for a number of particles: for x-rays, gammas and electrons a value of 22 eV with a spread of 1 eV has been measured \cite{vinagre, platzman, hurst, Ahlen:1980xr}, which is in good agreement with the 22--23~eV Monte Carlo calculated values for soft x-rays \cite{dias}. For alpha particles, consistent values of $20.9\pm 0.4$ eV \cite{Saito:2003dz,mimura} and 21.9 eV \cite{jesse} have been measured\footnote{For alpha particles, the energy loss mechanism takes place mainly through two steps. First, through the production of electrons (delta-rays). For a 6~MeV alpha particle colliding with a xenon atom, these delta-rays have a maximum energy of about 3~keV. Second, through the excitation and ionization of the neighbouring xenon atoms by these low-energy electrons.}. On the other hand, a much larger spread in the experimental determination of \Wex\ in high-pressure xenon gas exists, as shown in Tab.~\ref{tab:Wex}. The difference between these results is presently not fully understood. Given that for electrons with energies above about 200~eV the electron-xenon atom ionization cross-section is about 5--7 times larger than the excitation cross-section \cite{dias}, one expects \Wex$\gg$\Wi. For \Wi=22 eV and for the full range of \Wex\ values in Tab.~\ref{tab:Wex}, we obtain \Nex/\Ni\ values anywhere between 0.17 and 0.69. The ratio \Nex/\Ni\ has been directly measured before in similar experimental conditions as ours, namely high-pressure xenon gas and using alpha particles. The value \Nex/\Ni$=0.60\pm 0.03$ has been obtained for 1--10 bar gas pressures in \cite{Saito:2003dz}, while \cite{mimura} reports a consistent value of \Nex/\Ni$=0.61\pm 0.04$ at 5 bar. The gaseous xenon \Nex/\Ni\ values can be compared with those for liquid xenon, where our current understanding of energy partitioning is only marginally better. A \Nex/\Ni\ value of 0.06 has been estimated in this case \cite{Miyajima:1974zz}, although 0.20 is more consistent with experimental data \cite{Aprile:2007qd}. Their average, 0.13, is often used for liquid xenon for electron projectiles \cite{Chepel:2012sj}. 

The paper is organized as follows. The basic physics processes affecting the measurements of electron-ion recombination and energy partitioning between ionization and excitation in NEXT-DEMO are discussed in Sec.~\ref{sec:Recombination}. The data analysis is described in Sec.~\ref{sec:Analysis}, and results are given in Sec.~\ref{sec:Results}. We conclude in Sec.~\ref{sec:Conclusions}.

\section{Energy partitioning and electron-ion recombination in xenon} \label{sec:Recombination}

\begin{figure}[t!]
\begin{center}
\includegraphics[width=.70\textwidth]{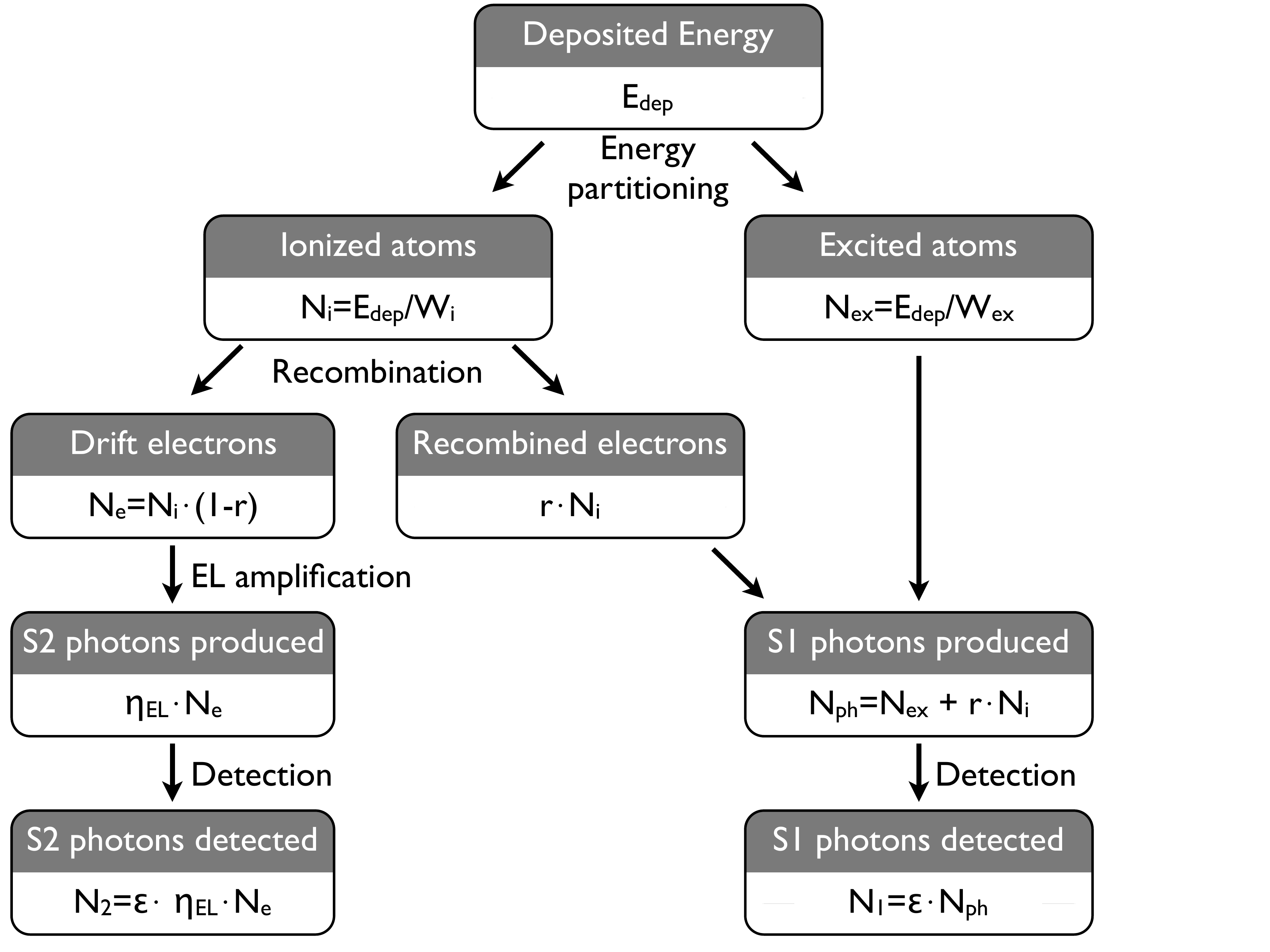} 
\end{center}
\caption{Processes affecting the detection of ionization and excitation in an electroluminescent TPC, together with the symbols defined in the text.}\label{fig:detctionprocess}
\end{figure}

The physical processes affecting the detection of ionization and scintillation in an {\it electroluminescent} xenon TPC are summarized in Fig.~\ref{fig:detctionprocess}. In order to describe the partitioning of energy loss processes between ionization and excitation, and to understand the absolute ionization and primary scintillation yields of xenon, it is customary to define three work functions \Wi, \Wex\ and \W\ as follows:
\begin{equation}
\Edep=\Wi\Ni=\Wex\Nex=\W(\Ni+\Nex)
\label{eq:edep}
\end{equation}
\noindent where \Edep\ is the deposited energy, \Wi\ (\Wex) is the average energy spent to ionize (excite) one xenon atom, and \W\ is the average energy to create {\it either} one electron-ion pair {\it or} one excited atom. In Eq.~\ref{eq:edep}, \Ni\ (\Nex) represents the number of ionized (excited) xenon atoms.

In the presence of electron-ion recombination, the number of electron-ion pairs \Ne\ produced by a ionizing track can be written as:
\begin{equation}
\Ne = \Ni(1-r)
\label{eq:ne}
\end{equation}
\noindent where $0\le r\le 1$ is the recombination probability. Given that each primary excitation \Nex\ and each recombined pair $r\cdot\Ni$ produces one scintillation photon \cite{Aprile:2009dv,Saito:2003dz}, the total number of scintillation photons \Nph\ can therefore be written as:
\begin{equation}
\Nph=\Nex+r\cdot\Ni
\label{eq:nph}
\end{equation}

From Eqs.~\ref{eq:ne} and \ref{eq:nph}, it immediately follows that the sum of electron and photon quanta produced is independent of recombination fluctuations, since $r$ cancels out in the sum:
\begin{equation}
\Ne+\Nph=\Ni+\Nex
\label{eq:neplusnph1}
\end{equation}

From Eqs.~\ref{eq:ne} and \ref{eq:nph}, it also follows that we can express the \Nex/\Ni\ ratio measuring the energy partitioning as:

\begin{equation}
\frac{\Nex}{\Ni} = (1-r)\cdot\frac{\Nph}{\Ne}-r
\label{eq:nexniratio1}
\end{equation}

For sufficiently high electric fields, we will see that $r\ll 1$, and the \Nex/\Ni\ ratio is essentially set by the light-to-charge ratio, \Nph/\Ne.

Furthermore, in the case of the electroluminescent TPC concept, both the ionization and scintillation signals are detected via the same array of photo-detectors, see Sec.~\ref{sec:Analysis}. The detected scintillation and ionization signals can be written as:
\begin{eqnarray}
\None(x,y,z)=\varepsilon_1(x,y,z)\cdot \Nph \label{eq:n1}\\
\Ntwo(x,y,z)=\varepsilon_2(x,y,z)\cdot \ELGain\cdot \Ne \label{eq:n2}
\end{eqnarray} 
In Eqs.~\ref{eq:n1} and \ref{eq:n2}, $N_1(x,y,z)$ and $N_2(x,y,z)$ are the so-called S1 and S2 detected signals, respectively, expressed in number of photoelectrons (PEs). The parameter \ELGain\ is the effective electron-to-photon amplification gain produced by the electroluminescent process, while the detector-dependent efficiencies $\varepsilon_1(x,y,z)$ and $\varepsilon_2(x,y,z)$ give the probability to detect a scintillation photon or an ionization electron produced at the detector position $(x,y,z)$, respectively. These efficiencies include photo-cathode coverage, quantum efficiency, light absorption and electron attachment effects, and therefore generally vary with spatial position $(x,y,z)$. In the special case of light and charge signals produced at zero drift distance ($z=0$) and along the cylindrical detector axis ($x=y=0$), these two efficiencies conveniently coincide for this type of detector, and will simply be written as $\varepsilon = \varepsilon_1(0,0,0)=\varepsilon_2(0,0,0)$ in the following. Unless otherwise noted, when we use the symbols \None\ and \Ntwo, without the explicit dependence on the spatial position, we refer to S1 and S2 signals that have been corrected to the spatial position $x=y=z=0$:
\begin{eqnarray}
\None=\Eff\cdot\Nph \label{eq:n1corr}\\
\Ntwo=\Eff\cdot\ELGain\cdot\Ne \label{eq:n2corr}
\end{eqnarray} 

From Eqs.~\ref{eq:neplusnph1}, \ref{eq:n1corr} and \ref{eq:n2corr}, the scintillation yield can be written as:
\begin{equation}
\None = \Eff\cdot(\Ni+\Nex)-\frac{\Ntwo}{\ELGain}
\label{eq:n1corr2}
\end{equation}

Equation \ref{eq:n1corr2} can be used to determine experimentally the value of the optical gain \ELGain, as we will see in Sec.~\ref{subsec:ResultsGain}. From Eqs.~\ref{eq:nexniratio1}, \ref{eq:n1corr} and \ref{eq:n2corr}, we can also express the \Nex/\Ni\ ratio as:

\begin{equation}
\frac{\Nex}{\Ni}=(1-r)\cdot\ELGain\cdot\frac{\None}{\Ntwo}-r
\label{eq:nexniratio2}
\end{equation}
\noindent which does not depend on the overall normalization for the light collection efficiency \Eff. In Eq.~\ref{eq:nexniratio2}, the recombination probability $r$ is intended as the value averaged over all events at a given drift field.

\section{Experimental setup} \label{sec:ExperimentalSetup}

\begin{figure}[t!]
\begin{center}
\includegraphics[width=.60\textwidth]{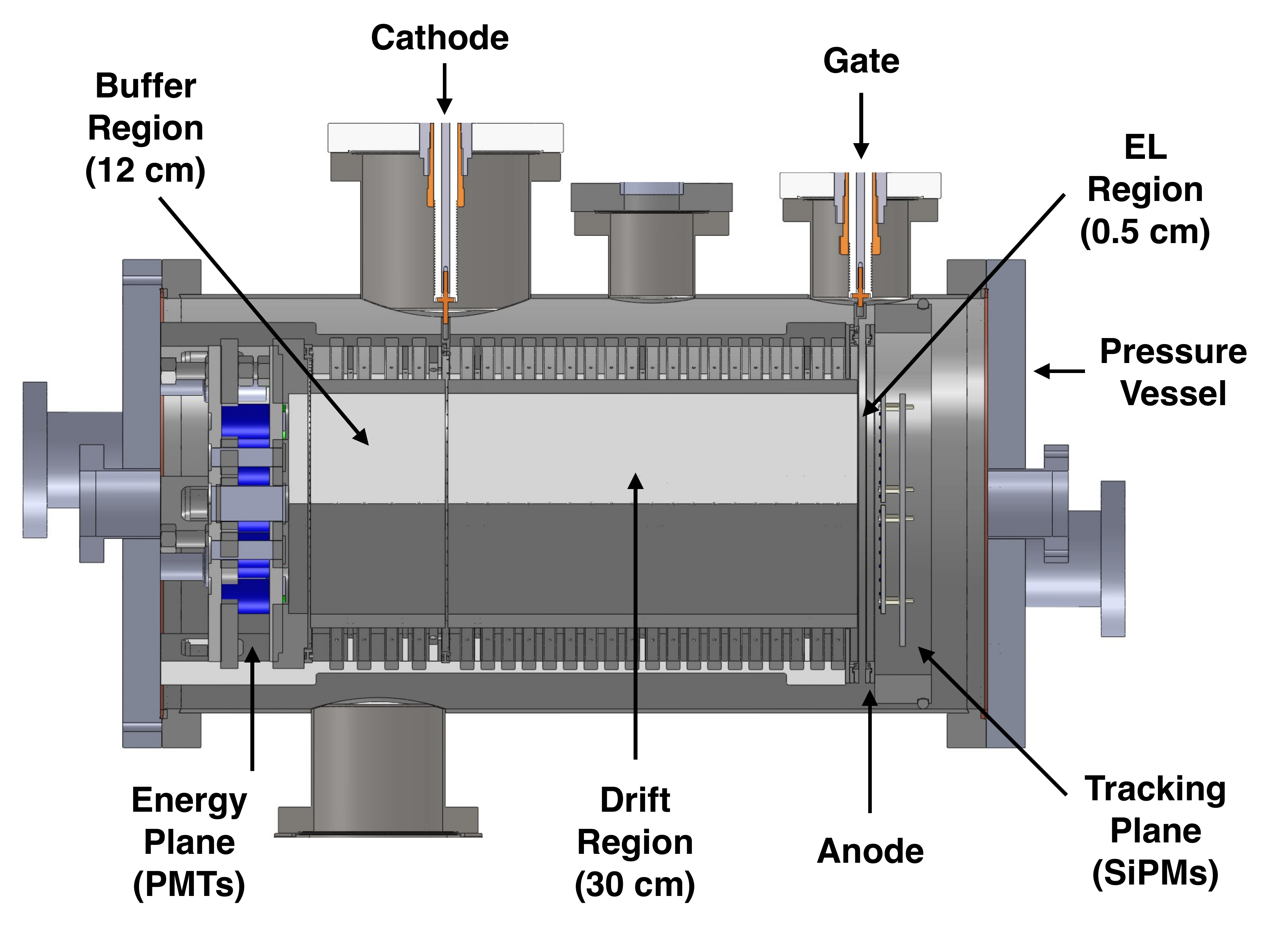}
\end{center}
\caption{Cross-sectional view of the NEXT-DEMO detector used in this analysis.}\label{fig:nextdemo}
\end{figure}

The NEXT-DEMO high-pressure xenon gas time projection chamber (TPC) is used in this analysis. A schematics of the detector is shown in Fig.~\ref{fig:nextdemo}. The TPC is filled with pure xenon gas at 10 bar pressure. The gas is continuously circulated through a room-temperature getter (model SAES MC50) to effectively remove impurities from the xenon. An important difference with respect to \cite{Alvarez:2012hu} is that, for this data set, the detector is instrumented with an additional tracking readout plane made of 256 silicon photomultipliers (SiPMs, model Hamamatsu S10362-11-050P). This readout plane complements the energy plane made of 19 photomultipliers (PMTs, model Hamamatsu R7378A). SiPMs are coated with a wavelength shifting molecule, tetraphenyl butadiene (TPB), in order to make the sensors sensitive to xenon scintillation light \cite{Alvarez:2012ub}. As mentioned in Sec.~\ref{sec:Recombination}, NEXT-DEMO is an electroluminescent (EL) TPC, where the ionization signal is converted into secondary (S2) scintillation light by a nearly noiseless amplification stage. Once the ionization electrons drift out of the active volume, they enter a 5~mm wide EL region defined by gate and anode meshes. The electric field in this region is sufficiently strong for electrons to excite, but not ionize, xenon atoms. Xenon de-excitation gives rise to the secondary scintillation light, with many photons produced per ionization electron. For further details about the NEXT-DEMO detector, see \cite{Alvarez:2013gxa}.

The calibration source is the same as in \cite{Alvarez:2012hu}. It is made of a dry radium ($^{226}$Ra) powder, it has an activity of 74 kBq and is connected to the gas system. The alpha decay of $^{226}$Ra produces $^{222}$Rn gas that diffuses through the entire detector. 

To avoid signal saturation for the energetic alpha events, NEXT-DEMO is operated in these runs with a low EL field, defined by a gate voltage of -5~kV and by a grounded anode. Such a field corresponds to an optical (EL) gain of order $10^2$ photons/electron only, see Sec.~\ref{subsec:ResultsGain}. Runs with different cathode voltages are used in this analysis, corresponding to drift fields in the 0.3--1 kV/cm range.

As a way to monitor gas purity, we have measured for each drift field setting how ionization electrons are exponentially suppressed as a function of drift time (see also Sec.~\ref{subsec:AnalysisCorrections}). The charge attenuation is due to electron attachment to electronegative impurities in the xenon gas, such as O$_2$. The measured electron lifetimes, $(3.50\pm 0.15)$~ms for all drift field settings, are much longer than the typical detector drift times (of order $\mathcal{O}(100~\mu$s)), indicative of high gas purity and small attachment effects. In particular, assuming that impurities are entirely due to O$_2$, the measured lifetimes correspond to 100 parts-per-billion or less of O$_2$ \cite{Biagi:1999nwa}.
\section{Analysis} \label{sec:Analysis}

\subsection{Selection of alpha candidate events} \label{subsec:AnalysisSelection}

The first selection of alpha candidate events uses the S2 and S1 signals as recorded by the PMTs in the energy plane, as done in \cite{Alvarez:2012hu}. We require events to have a single S2-like peak with $10^3$ photoelectrons (PEs) minimum charge in each individual PMT channel and with a time width not exceeding 30~$\mu$s. The width requirement is introduced to isolate alpha-like events from other energetic events, such as cosmic ray muons. We require events to have a single S1-like peak with 20 PEs minimum charge in each PMT and a peak risetime of less than 0.18~$\mu$s.  

\begin{figure}[t!]
\begin{center}
\includegraphics[width=.45\textwidth]{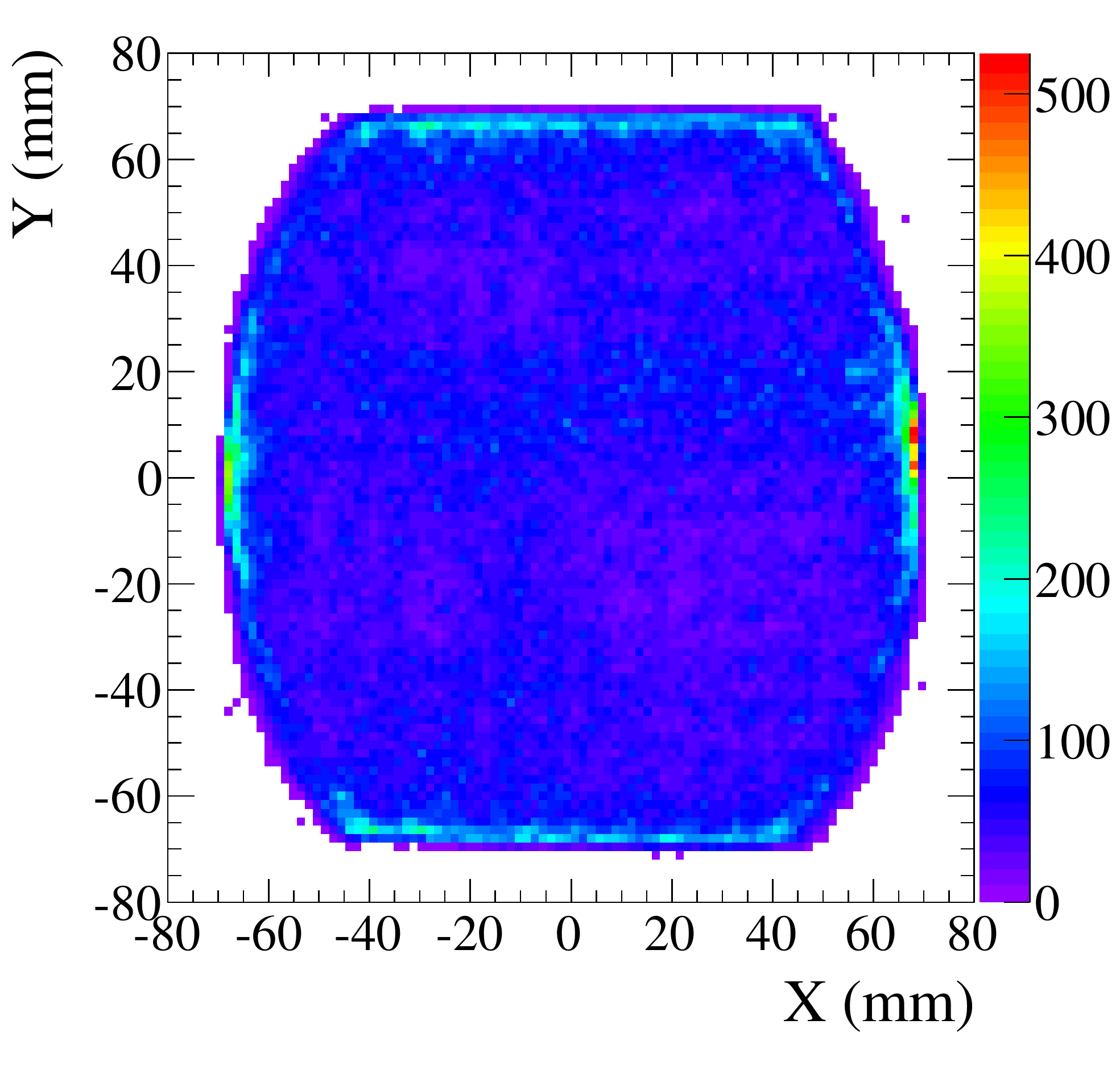} \hfill
\includegraphics[width=.45\textwidth]{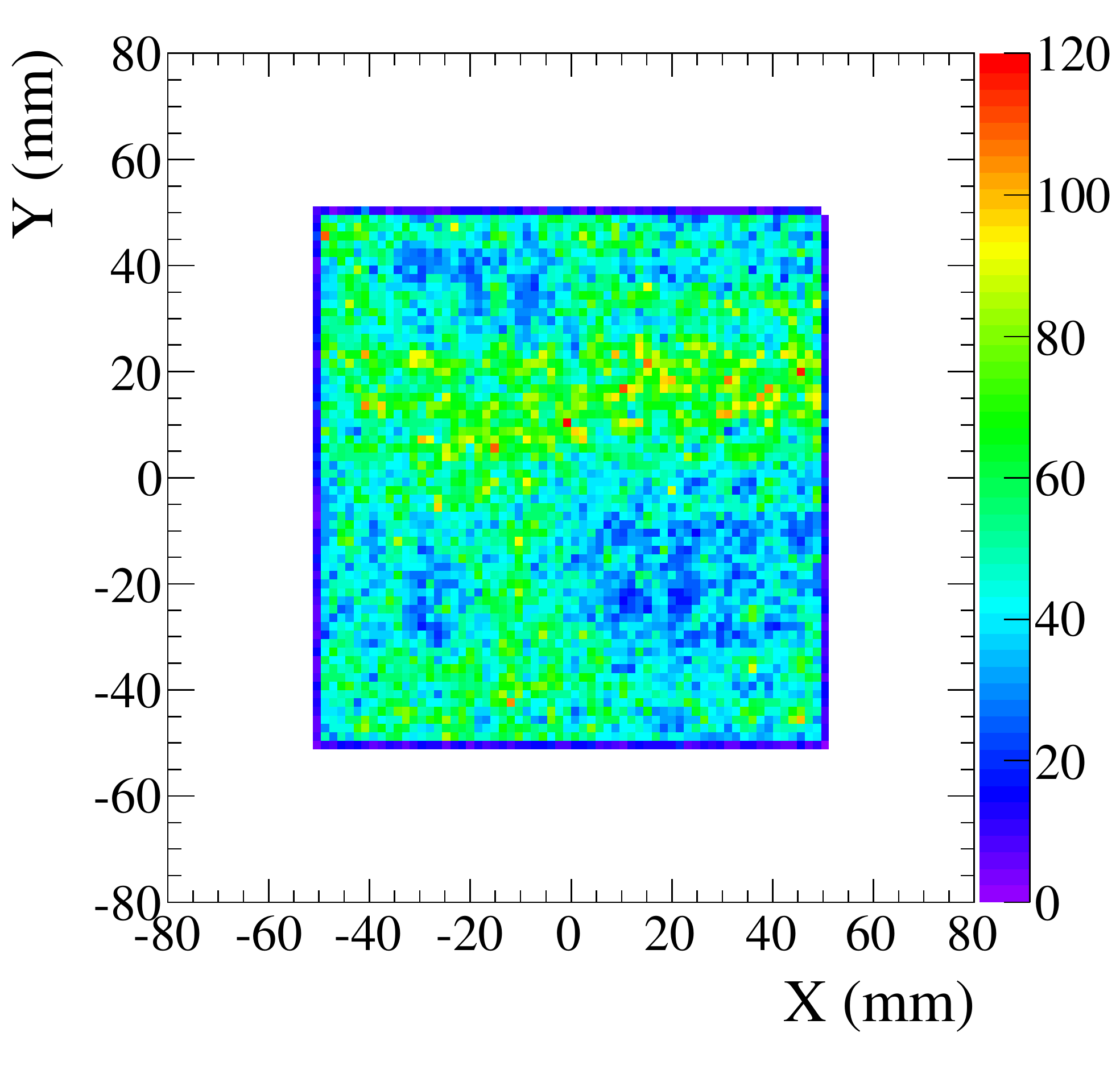}
\end{center}
\caption{Spatial distribution of alpha candidate events in the $(x,y)$ plane. The $x,y$ positions are reconstructed from the SiPM tracking plane using a barycenter algorithm. Left panel: all events. Right panel: selected events, requiring $\lvert x,y\rvert<50$~mm. }\label{fig:yvsx}
\end{figure}

The next step in the selection uses S2 information as recorded from the SiPMs in the tracking plane, and it is new with respect to \cite{Alvarez:2012hu}. We require events to have a maximum of 50 (out of 248) SiPM hits, and that the reconstructed $x$ and $y$ positions of the events satisfy $\lvert x,y\rvert<50$ mm, where $(x,y)$ defines the plane perpendicular to the drift direction and $x=y=0$ is the detector central axis. The event $x$ and $y$ positions have been reconstructed using a simple barycenter algorithm described in \cite{Alvarez:2013gxa}, appropriate given that alpha energy depositions in 10 bar pressure xenon gas are nearly point-like. Figure \ref{fig:yvsx} shows the spatial distribution of alpha candidate events before (left panel) and after (right panel) the $\lvert x,y\rvert<50$ mm cut. This requirement is needed to ensure full energy containment in the detector active volume, and improves drastically the energy resolution of the detector compared to \cite{Alvarez:2012hu}. The distribution of events before the $x,y$ cut is reminiscent of the hexagonal shape of the detector cross-sectional active area, distorted by the incomplete coverage with SiPMs at large $\lvert  x\rvert$ values. The spatial distribution of events is not uniform after the $x,y$ cut, either. On the one hand, 8 out of 248 SiPMs (3.2\%) were not functional during this run, causing small-scale inhomogeneities in reconstructed positions. On the other hand, larger-scale inhomogeneities may be caused by a non-uniform gas flow inside the detector. The final selection requirement is on the drift length $z$ of alpha candidate events, required to be comprised between 40 and 290~mm. Compared to \cite{Alvarez:2012hu}, we extend the fiducial volume fraction along $z$ from 60\% to 83\% of the full, 300~mm long, detector drift length. 


\subsection{Spatial corrections to ionization and scintillation signals} \label{subsec:AnalysisCorrections}

Both ionization (\Ntwo) and scintillation (\None) yields recorded on the energy plane are affected by instrumental effects, causing the observed number of photons to depend on the alpha particle spatial position $(x,y,z)$ within the detector. In this Section, we show the spatial dependence of the yields on the alpha candidate events defined in Sec.~\ref{subsec:AnalysisSelection}, and we describe our corresponding correction procedure. As mentioned in Sec.~\ref{sec:Recombination}, the yields for the events in the $(-50<x\rm{ (mm)}<50, -50<y\rm{ (mm)}<50, 40<z\rm{ (mm)}<290)$ fiducial volume are corrected to yields at the $x=y=z=0$ position.

\begin{figure}[t!]
\begin{center}
\raisebox{-0.5\height}{\includegraphics[width=.50\textwidth]{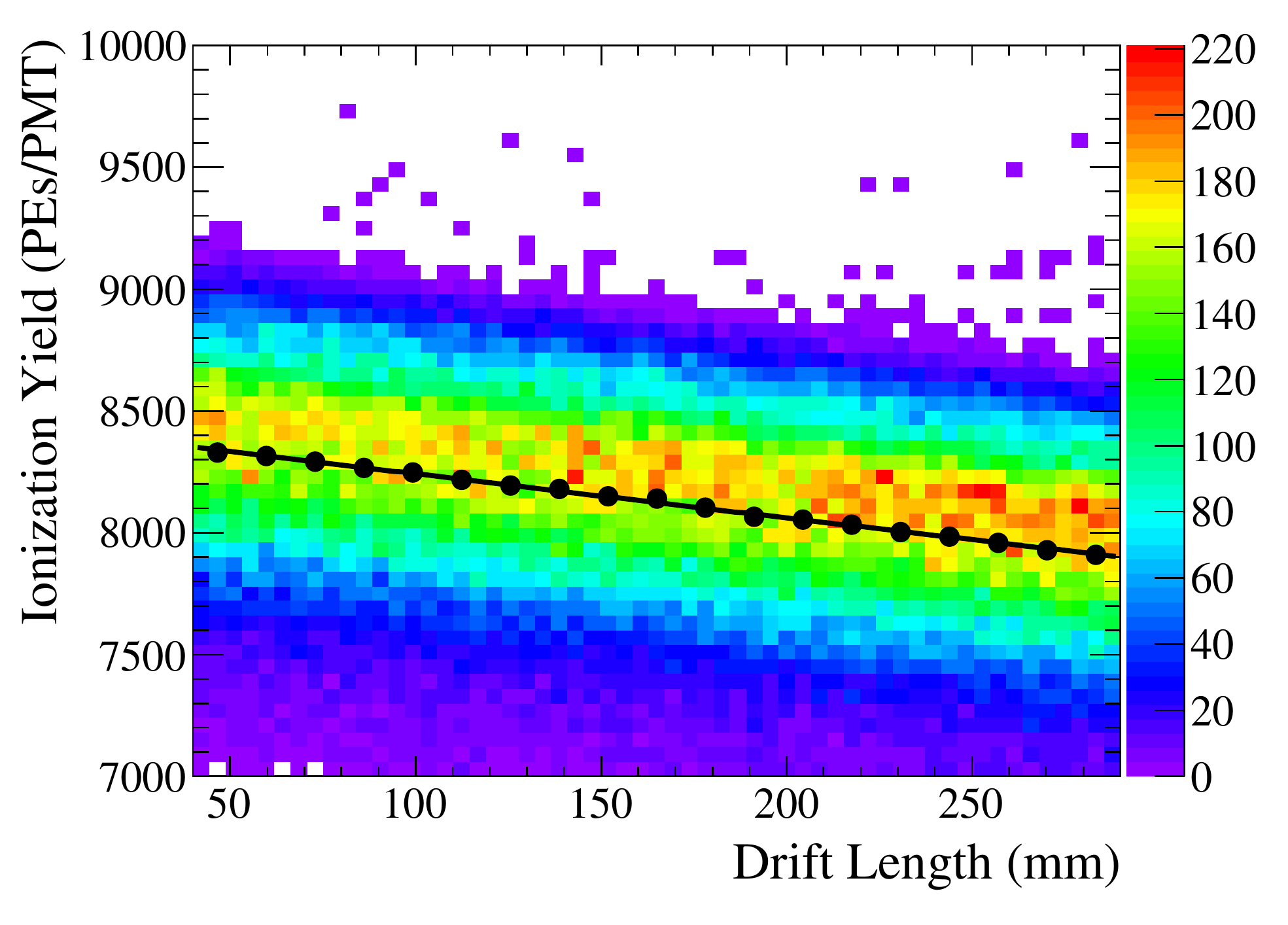}} \hfill
\raisebox{-0.5\height}{\includegraphics[width=.45\textwidth]{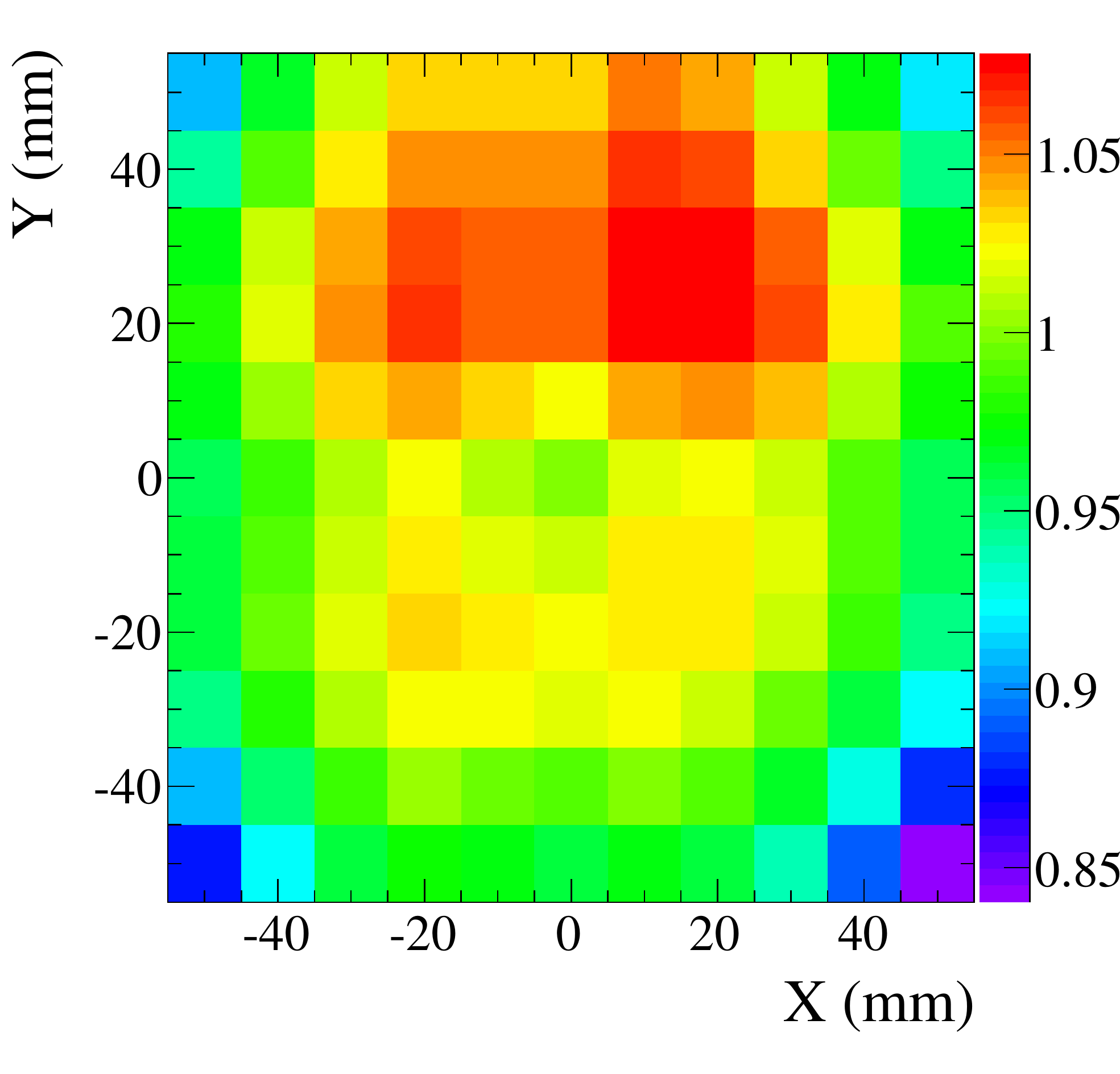}}
\end{center}
\caption{Ionization (S2) yield of alpha candidate events as a function of spatial position. Left panel: yield (in PEs/PMT) versus drift length ($z$-position). Data are fitted to an exponential function, shown by the black curve and accounting for electron attachment. Right panel: yield versus $(x,y)$ position, computed in 10~mm$\times$10~mm wide bins. The yields are normalized to the response on the detector central axis ($x=y=0$).}\label{fig:s2spatialcorr}
\end{figure}

The spatial dependence of the ionization yield \Ntwo\ can naturally be split into a $z$-dependent variation due to electron attachment along drift, and a separate $(x,y)$-dependent effect. The latter dependence is due to the variation in EL light response with $(x,y)$ introduced by possible non-uniformities in the EL gap width, and by the non-perfect reflectivity of the light tube. The same pattern in $(x,y)$ for \Ntwo\ light in NEXT-DEMO has also been obtained using xenon X-rays \cite{Lorca:2014sra}. The $z$ and $(x,y)$ dependencies are shown in the left and right panel of Fig.~\ref{fig:s2spatialcorr}, respectively. As expected, the $z$ dependence is well described by an exponential function, shown in the left panel of Fig.~\ref{fig:s2spatialcorr} and used to correct the yield. This correction does not exceed the 5\% level over the full drift length. The correction in $(x,y)$ is purely empirical, and done bin-by-bin in $\Delta x=\Delta y=10$ mm wide bins. As shown in the right panel of Fig.~\ref{fig:s2spatialcorr}, the \Ntwo\ yields vary in the 0.84--1.08 range relative to the one at $x=y=0$ used as reference, over the $-50<x,y<50$ mm region.

\begin{figure}[t!]
\begin{center}
\raisebox{-0.5\height}{\includegraphics[width=.50\textwidth]{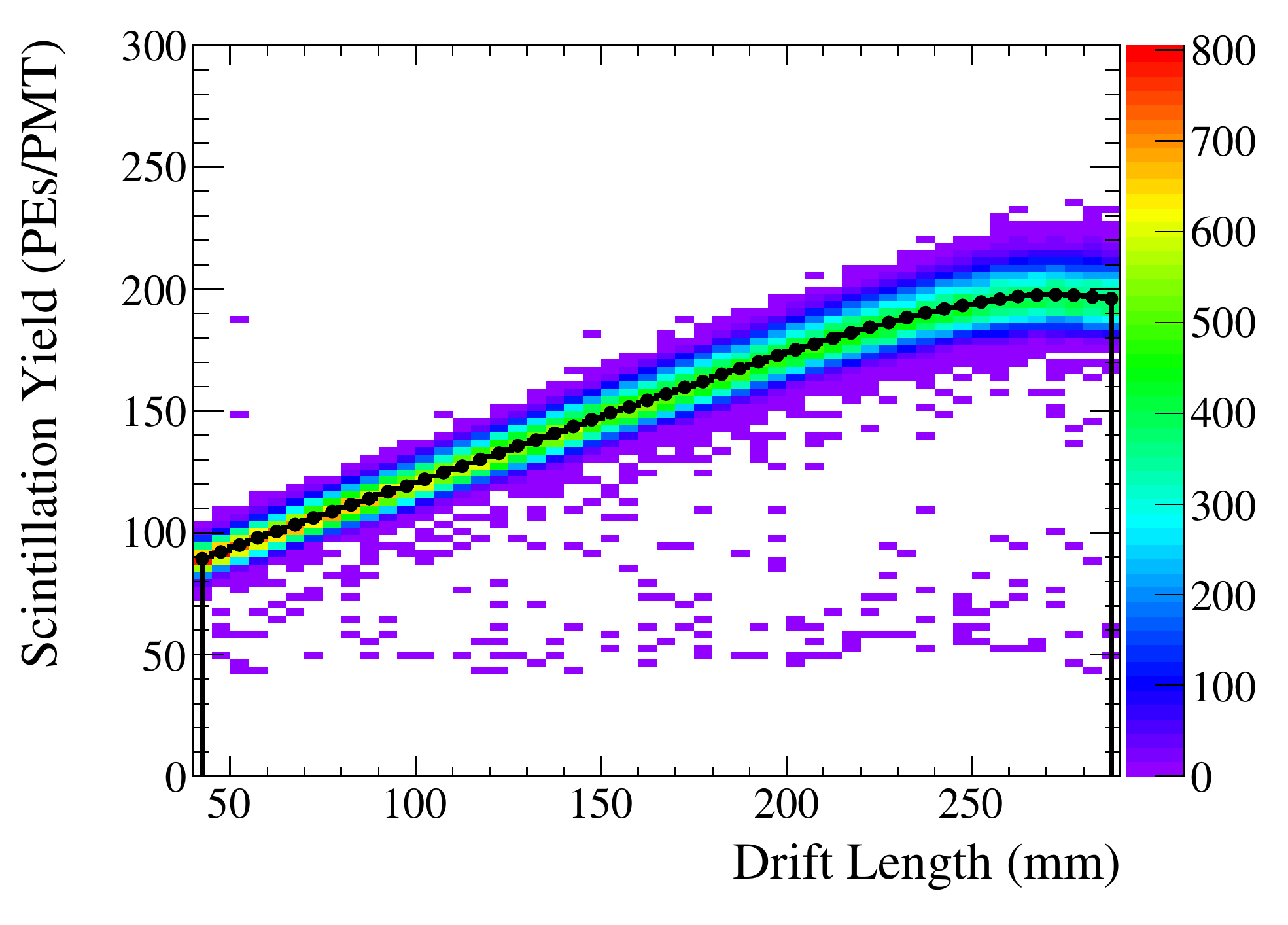}} \hfill
\raisebox{-0.5\height}{\includegraphics[width=.45\textwidth]{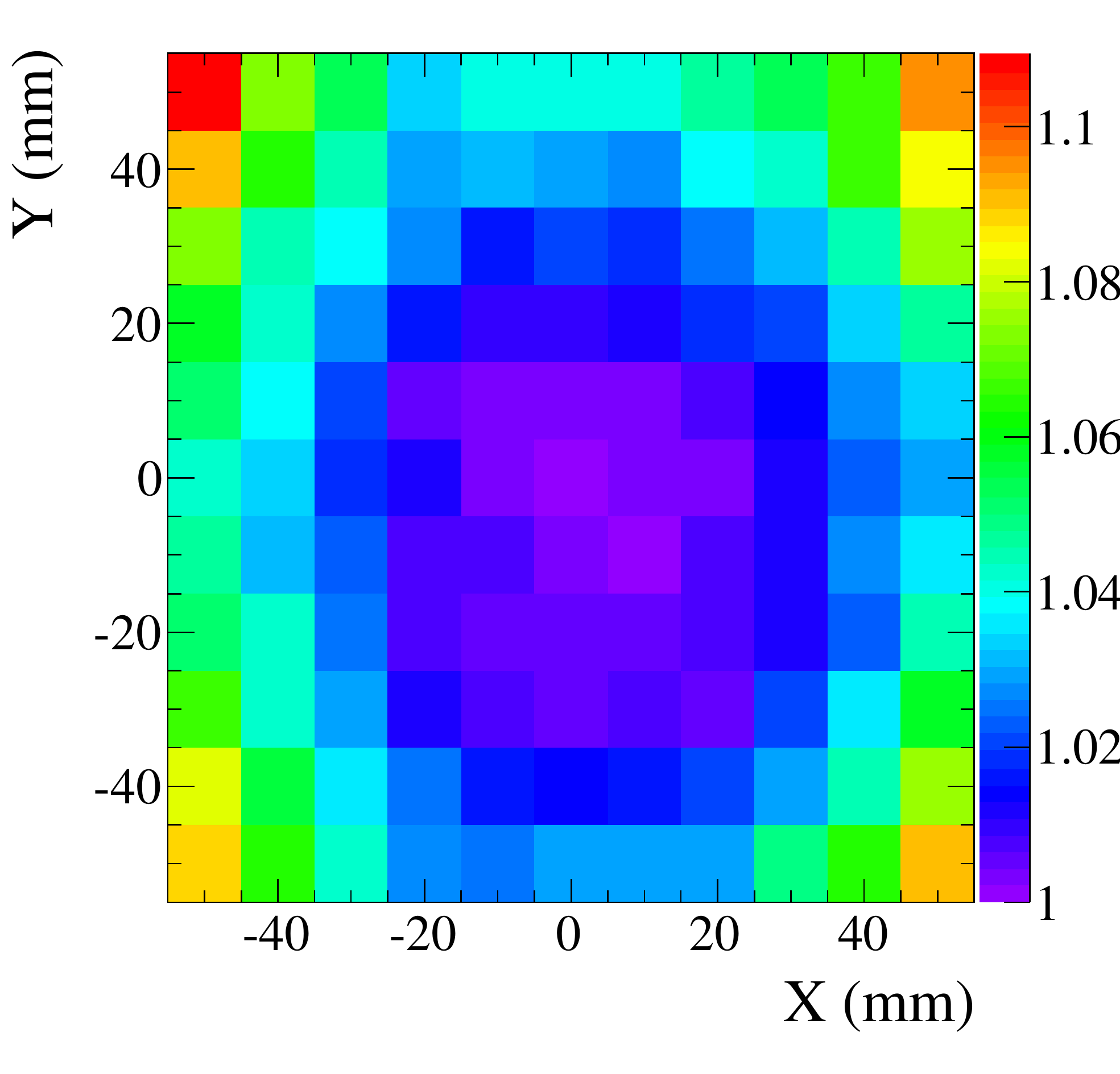}}
\end{center}
\caption{Scintillation yield \None\ of alpha candidate events as a function of spatial position. Left panel: yield (in PEs/PMT) versus drift length ($z$ position). The black curve shows the empirical spline fit to the data. Right panel: relative yield versus $(x,y)$ position. The correction procedure in this case is identical to the one in Fig.~\protect\ref{fig:s2spatialcorr}, except that it is done separately for five $z$ regions ($240<z<290$~mm region shown here).}\label{fig:s1spatialcorr}
\end{figure}

On the other hand, a full 3-dimensional spatial correction to the scintillation yield \None\ is in principle necessary, given that $z$ and $(x,y)$ corrections are correlated with each other in this case. An approximation is made here for simplicity, motivated by the fact that the $z$ dependence is the dominant one. As discussed in \cite{Alvarez:2012hu}, the scintillation yield per unit energy deposition is lower near the EL region (low $z$ values) primarily because of the non-perfect optical transparency of the gate and anode meshes, with an open area of 76\% and 88\%, respectively. The non-perfect reflectivity of the light tube, particularly in the uncoated buffer region between the cathode and the energy plane, also contributes to this $z$ dependence. As a result, variations of order 100\% in \None\ for alpha candidate events in the $40<z<290$~mm region can be seen in the left panel of Fig.~\ref{fig:s1spatialcorr}. An emprirical spline fit to the data (also shown in Fig.~\ref{fig:s1spatialcorr}) is used to correct the \None\ yield to $z=0$. Once the $z$ correction is done, the sub-dominant $(x,y)$ correction on \None\ is made. In this case, the same procedure described above for \Ntwo\ is done, with the exception that it is done separately for five coarse, 50~mm wide, $z$ regions between 40 and 290~mm. The $(x,y)$ dependence of the \None\ yield in one of these $z$ regions ($240<z<290$~mm) is shown in the right panel of Fig.~\ref{fig:s1spatialcorr}, with variations of order 10\% as for \Ntwo. The inhomogeneities in the \None\ $(x,y)$ pattern are found to decrease for smaller $z$ values (further away from the energy plane), with inhomegenities at the 3\% level only for $40<z<90$~mm. The $\Delta x=\Delta y=10$~mm and $\Delta z=50$~mm binning chosen provides enough granularity as well as sufficient statistical precision (at least $10^3$ events per bin) for the \None\ yield correction.

As we will see in Sec.~\ref{sec:Results}, the majority of the selected alpha candidate events are due to \Rn{222}\ (5.49 MeV) decays, with a small (15--23\%, depending on the drift field setting) fraction due to higher-energy alphas. In order to minimize biases in our calibration procedure shown in Figs.~\ref{fig:s2spatialcorr} and \ref{fig:s1spatialcorr}, only mono-energetic \Rn{222}\ events are used to correct for the spatial dependence of S1 and S2 yields. This is done by imposing a $E<5.80$~MeV energy cut on the alpha candidates sample. Since our energy estimator $E$ depends in turn on such spatial corrections, an iterative procedure is used. Two iterations are found to be sufficient.

\section{Results} \label{sec:Results}

\subsection{Correlated fluctuations between ionization and scintillation} \label{subsec:ResultsCorrelatedFluctuations}

The simultaneous measurement of scintillation and ionization allows the study of correlated fluctuations among them. Because of the small Fano factor of xenon gas (for $\rho < 0.5$~g/cm$^{3}$ densities), fluctuations in the ratio of excited to ionized xenon atoms are expected to be negligible, at the $10^{-3}$ level. For this reason, once detector calibration is performed, all anti-correlated fluctuations between scintillation and ionization can be interpreted as being caused by fluctuations in the electron-ion recombination process. In practice, the measured S1-S2 correlation coefficients are expected to be larger than -1 even for a mono-energetic alpha sample, because of instrumental effects and statistical fluctuations in the S1 and S2 detection process. For a discussion of various effects contributing to the measured S1-S2 correlation coefficient, see Ref.~\cite{Aprile:2007qd}.

\begin{figure}[t!]
\begin{center}
\includegraphics[width=.48\textwidth]{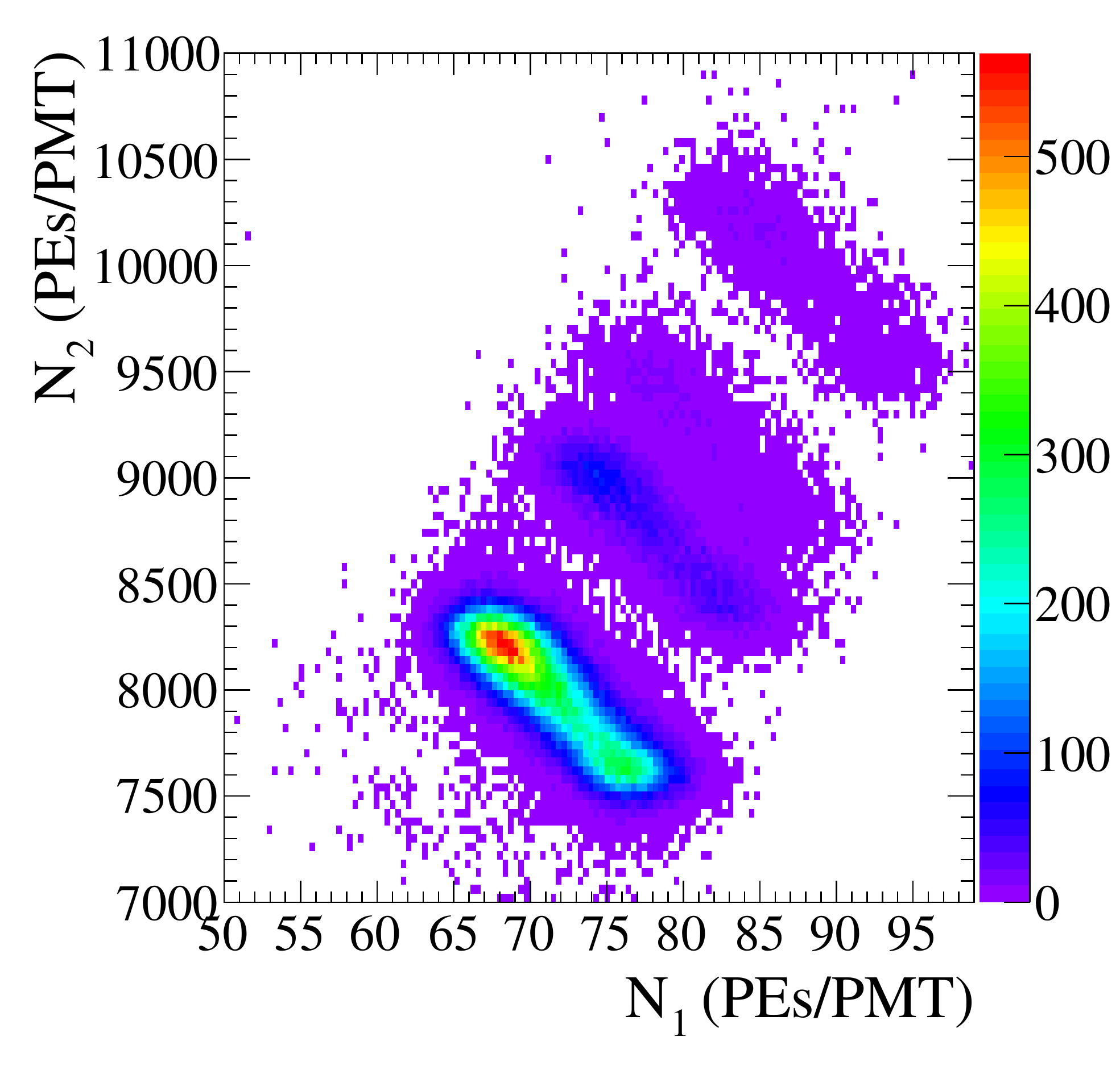} \hfill
\includegraphics[width=.48\textwidth]{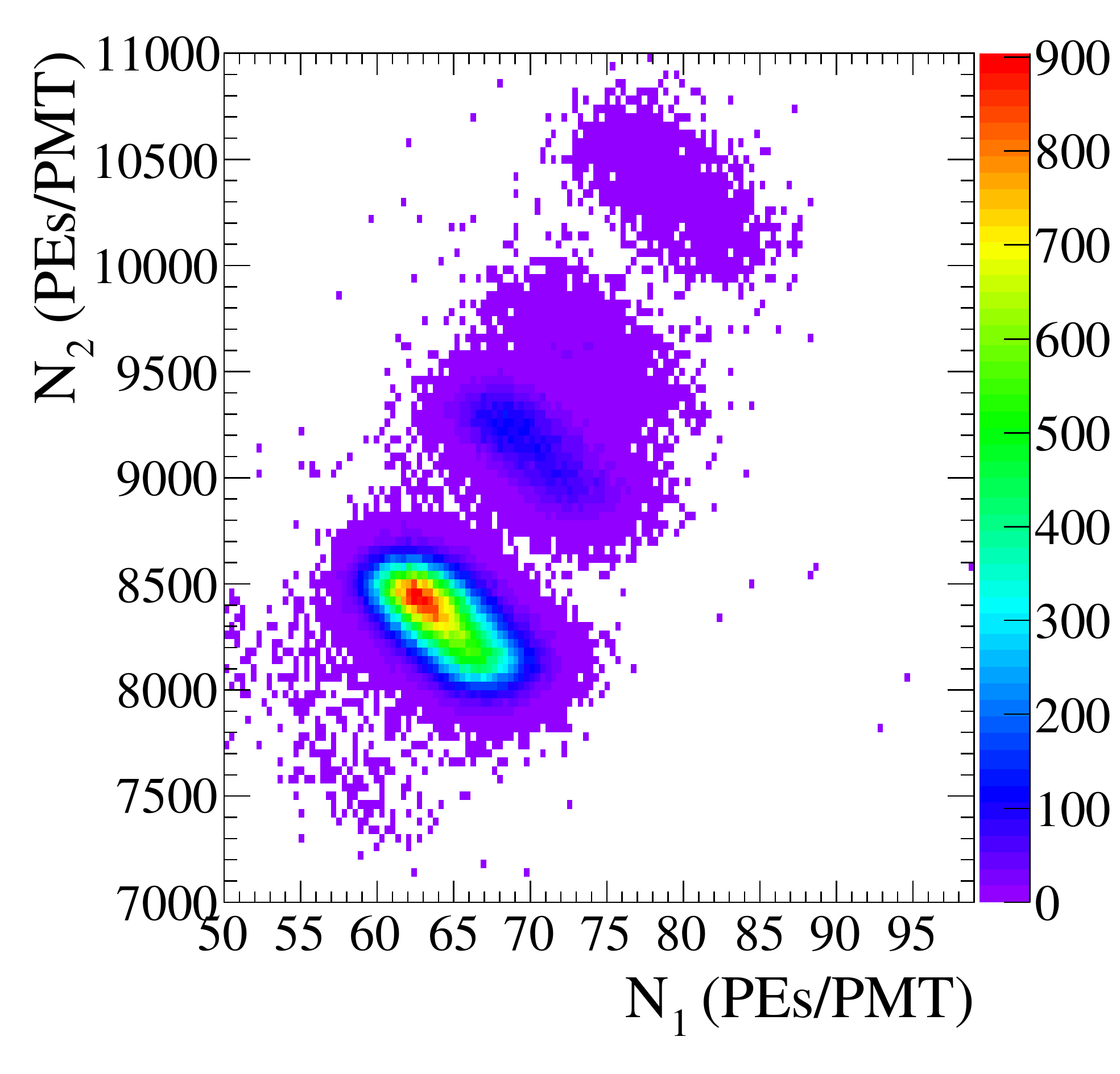}
\end{center}
\caption{Ionization signal (\Ntwo) versus primary scintillation signal (\None) for drift field of 0.3~kV/cm (left) and 1~kV/cm (right). Four alpha decays can be seen, from lower to higher energies: \Rn{222}, \Po{218}, \Rn{220}, \Po{216}.}\label{fig:s2vss1}
\end{figure}

Figure~\ref{fig:s2vss1} shows the ionization (\Ntwo) as a function of the scintillation (\None) yields, for alpha candidate events collected in the 0.3~kV/cm and 1~kV/cm field configurations. Both yields have been corrected for spatial effects described in Sec.~\ref{subsec:AnalysisCorrections}. Four alpha decay populations can be identified in the figure. Most events correspond to alpha decays from \Rn{222}\ (5.49 MeV). The second most abundant population is due to \Po{218} (6.00 MeV). Both event samples are primarily produced by the radium source, see Sec.~\ref{sec:ExperimentalSetup}, as in \cite{Alvarez:2012hu}. The anti-correlation between \Ntwo\ and \None\ for alpha candidate events is clearly visible at all drift fields, with correlation coefficients between -0.80 (at 0.3~kV/cm) and -0.56 (at 1~kV/cm) for \Rn{222}-only events. As expected, the anti-correlation between scintillation and ionization is the strongest at the lowest drift values, where recombination fluctuations are the most pronounced. These correlation coefficients can be compared with the -0.29 (0.3~kV/cm) and -0.21 (1~kV/cm) values obtained in \cite{Alvarez:2012hu} for the same drift field strengths, and for the same gas pressure (10~bar) and for the same radioactive source (\Rn{222}). This much improved anti-correlation measurement is primarily due to the use of tracking information in the present analysis. As described in Sec.~\ref{subsec:AnalysisSelection}, we have now been able to eliminate events with incomplete energy depositions inside the fiducial volume, which caused a low energy tail in the \Ntwo\ distribution, as well as artificially large fluctuations, in our previous work \cite{Alvarez:2012hu}. This improved analysis also allowed us to identify, for the first time in the NEXT-DEMO detector, two additional alpha decay populations in Fig.~\ref{fig:s2vss1}: \Rn{220}\ (6.29 MeV) and \Po{216}\ (6.78 MeV). The latter alpha decays are not caused by the \Ra{226}\ source, but are due to radioactive impurities present in the xenon active volume.

\subsection{Optical gain} \label{subsec:ResultsGain}

The anti-correlation between S1 and S2 yields in the presence of recombination is expected to follow a linear dependence, considering that each recombined electron produces one additional scintillation photon. Mathematically, this can be seen in Eq.~\ref{eq:n1corr2}. In the presence of recombination fluctuations only, with no instrumental effects nor Poisson fluctuations in the number of detected photons, \None depends linearly on \Ntwo, with a slope set by the inverse of the optical gain, \ELGain. This relationship can be used to provide a measurement of \ELGain\ in an electroluminescent TPC.

\begin{figure}[t!]
\begin{center}
\includegraphics[width=.50\textwidth]{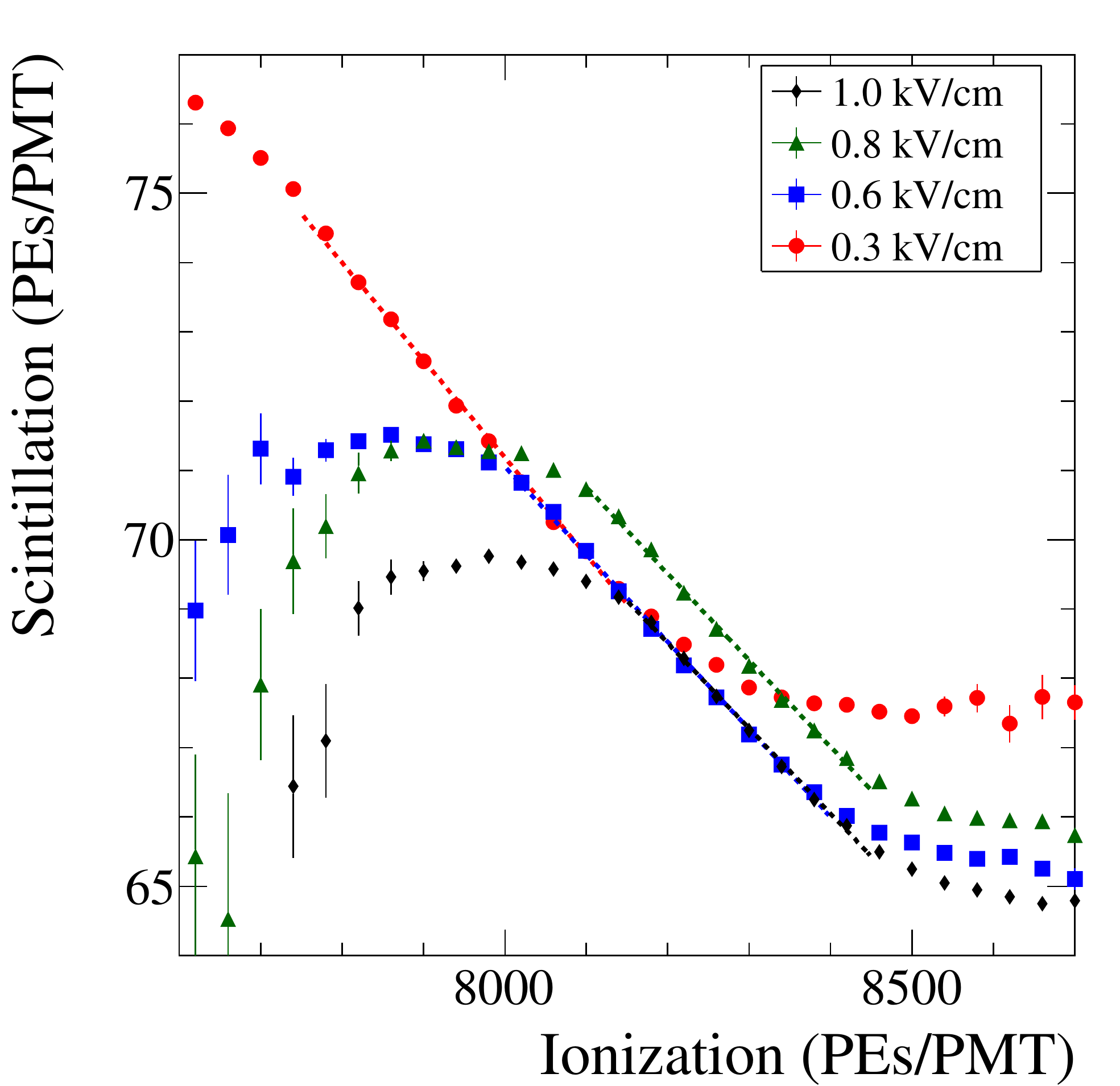} \hfill
\end{center}
\caption{Profile histogram of scintillation versus ionization yield for \Rn{222}\ events, for the four drift field settings used in this analysis. The central portion of the profile histograms exhibit a linear behavior shown by the dashed lines, with a constant slope set by the optical gain of the detector.}\label{fig:s1s2profile}
\end{figure}

Figure \ref{fig:s1s2profile} shows the the profile histogram of the scintillation yield \None, binned as a function of the ionization yield \Ntwo, for the four drift field settings used in this analysis and for \Rn{222}\ events only. In other words, we bin along the y-axis the \Rn{222}\ events in the 2D histograms of Fig.~\ref{fig:s2vss1}. We choose to bin along \Ntwo\ as opposed to \None\ because the ionization yield is the least affected among the two by Poisson fluctuations in the detection process, see Sec.~\ref{subsec:ResultsResolutionSpectroscopy}. The profile histograms exhibit a linear dependence between two ``plateaus'' at low and high \Ntwo\ values. The position and range of this linear range changes with the drift field settings: it is centered at lower \Ntwo\ values, and covers a wider \Ntwo\ range, for lower drift field settings, where recombination fluctuations are more important. However, the slope of this linear range is unaffected by the drift field. A linear fit to the profile histograms (also shown in Fig.~\ref{fig:s1s2profile}) yields:
\begin{equation}
\ELGain = (75\pm 5)
\label{eq:elgain1}
\end{equation}
\noindent where the uncertainty covers differences between drift field settings and reasonable variations in the fit range in \Ntwo. Outside of the linear range, Poisson fluctuations in the number of detected S2 photons affect significantly the profile histograms, and we therefore exclude the two extremes from the fit.    

This value can be compared with the expected EL gain at our xenon gas density and EL field conditions \cite{Monteiro:2007vz}:
\begin{equation}
\ELGain = \mathcal{T}\cdot [0.140\frac{E}{N}-0.474]N\cdot \Delta z = \mathcal{T}\cdot (129\pm 36) = (87\pm 25)
\label{eq:elgain2}
\end{equation}
\noindent where the EL region width $\Delta z=(0.47\pm 0.03)$ is expressed in cm, the xenon gas density $N=2.56\cdot 10^3$ in $10^{17}$ atoms per cm$^3$ (corresponding to our 9.8 bar operating pressure), the reduced field $E/N=4.15$ in $10^{-17}$V$\cdot$cm$^2$, and $\mathcal{T}$ is the dimensionless gate transparency. Optical simulations of light isotropically produced on both sides of a gate mesh with 76\% open area (as in NEXT-DEMO), and detected in the NEXT-DEMO energy plane, yield an expected gate transparency of $\mathcal{T}$=0.68. The uncertainty in the EL width (0.03 cm) is estimated from the difference between the surveyed (0.47 cm) and the nominal (0.50 cm) value, and is responsible for the large \ELGain\ uncertainty quoted in Eq.~\ref{eq:elgain2}. The measured value (Eq.~\ref{eq:elgain1}) is therefore in excellent agreement with the value predicted by the parametrization reported in Eq.~\ref{eq:elgain2} and \cite{Monteiro:2007vz}. We note that a number of linear parametrizations of the type of Eq.~\ref{eq:elgain2} exist in the literature, and that they can yield quite different predictions for \ELGain\ near EL threshold operations as adopted in this work.

\subsection{Energy reconstruction} \label{subsec:ResultsResolutionSpectroscopy}

As discussed in Sec.~\ref{sec:Recombination}, the sum of ionization electrons escaping recombination, \Ne=\Ntwo/($\varepsilon\cdot$\ELGain), plus scintillation photons, \Nph=\None/$\varepsilon$, is in principle free of recombination fluctuations. The event energy $E$ is defined as:
\begin{equation}
E = \lambda(\None+\Ntwo/\ELGain)
\label{eq:energy}
\end{equation}
\noindent where \ELGain\ has been determined experimentally in Sec.~\ref{subsec:ResultsGain}, and the energy scale factor $\lambda$ is obtained by aligning \Rn{222}\ events with the 5.49~MeV energy.

\begin{figure}[t!]
\begin{center}
\includegraphics[width=.65\textwidth]{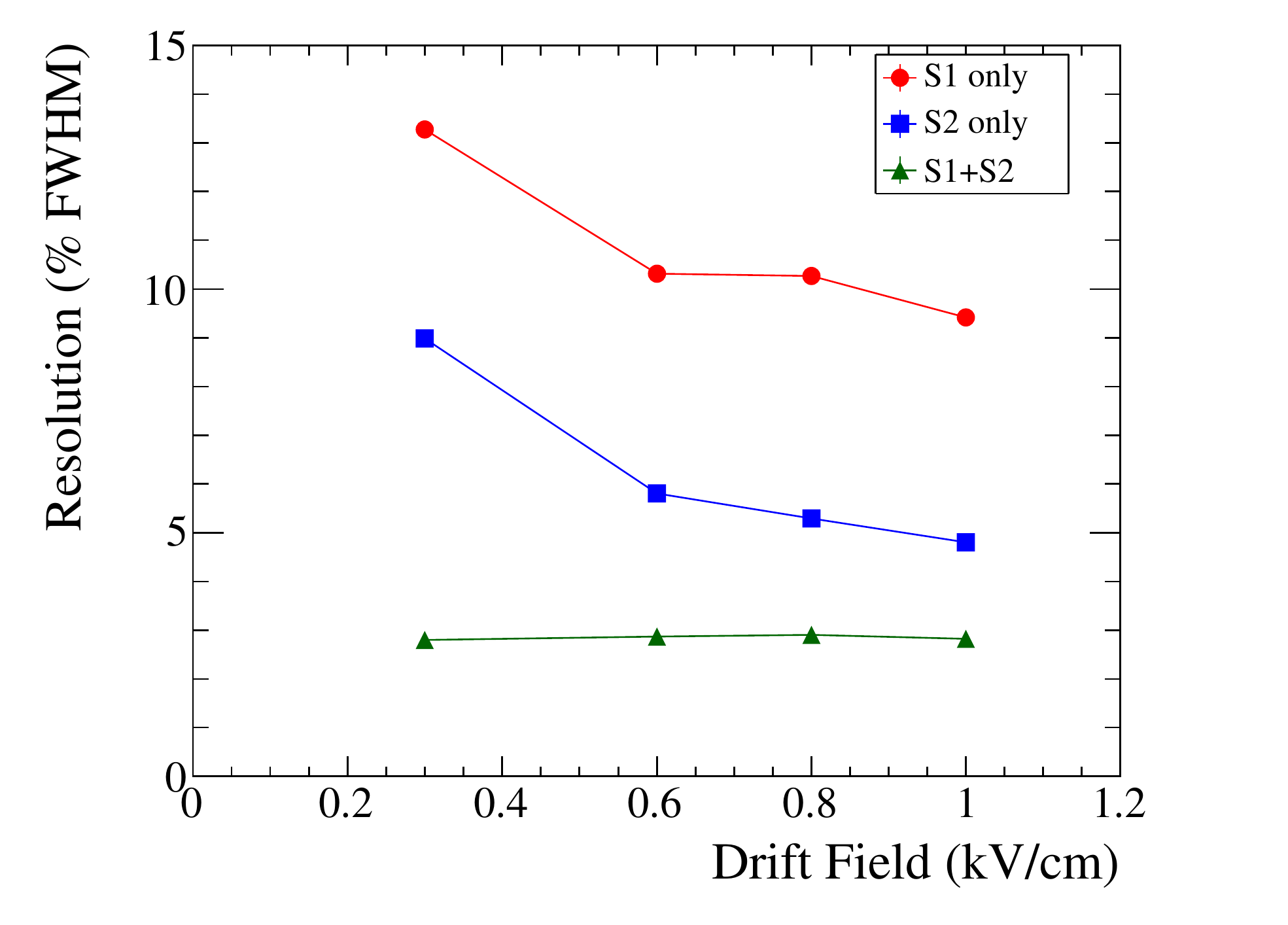} 
\end{center}
\caption{Energy resolution for \Rn{222}\ events as a function of drift field, using the S1 signal only, S2 only, or a combination of the two given by Eq.~\protect\ref{eq:energy}.}\label{fig:energyresolution}
\end{figure}

Figure~\ref{fig:energyresolution} shows the energy FWHM resolution of \Rn{222}\ events as a function of drift field, for three different energy estimators: \None\ as in Eq.~\ref{eq:n1corr} (S1 only), \Ntwo\ as in Eq.~\ref{eq:n2corr} (S2 only), and $E$ as in Eq.~\ref{eq:energy} (S1 plus S2). The best energy estimator is given by $E$ in Eq.~\ref{eq:energy}, with a FWHM resolution of about 2.8\% at 5.49~MeV. In addition, this resolution is independent of drift field values, and fluctuations are gaussianly-distributed. The \None\ and \Ntwo\ energy estimators have a worse resolution dominated by recombination fluctuations, particularly at low drift fields where such fluctuations are larger. Also, fluctuations have a non-gaussian behavior in this case. The S2-only energy estimator is better than the S1-only one, because of the nearly noiseless amplification stage provided by electroluminescence. This 2.8\% FWHM energy resolution is about a factor of 3 improvement over our previous result obtained with the same alpha calibration source in \cite{Alvarez:2012hu}, and in similar gas pressure and drift field conditions.

The measured energy resolution is in reasonable agreement with the ideal one expected for no recombination fluctuations. In our case, fluctuations in the number of excited and ionized xenon atoms produced can be neglected, as they are small compared to the Poisson fluctuations in the S2 and (especially) S1 signal detection process. The expected FWHM resolution is therefore:
\begin{equation}
\frac{\delta E}{E}\simeq2.354\frac{\sqrt{1+(\sigma_q/\overline{q})^2}}{\None+\Ntwo/\ELGain}
\sqrt{\delta\None^2+(\delta\Ntwo/\ELGain)^2}
\label{eq:idealresolution}
\end{equation}
\noindent where $\sigma_q/\overline{q}\simeq 1$ (see \cite{Lorca:2014sra}) is the typical single-PE charge resolution for NEXT-DEMO PMTs, $\ELGain\simeq 75$ (see Sec.~\ref{subsec:ResultsGain}), $\delta\None=\None/\sqrt{N_{1d}}$ and $N_{1d}$ is the number of detected S1 photons, and similarly for $\delta\Ntwo$. For \Rn{222}\ events in the 1~kV/cm drift field configuration, \None=66.92~PEs/PMT and \Ntwo=8332~PEs/PMT (see Fig.~\ref{fig:s2vss1}), $N_{1d}=153.2\cdot 19$ and $N_{2d}=8122\cdot 19$ (see Figs.~\ref{fig:s2spatialcorr} and \ref{fig:s1spatialcorr}, where 19 is the number of PMTs in the enegy plane), resulting in $\delta E/E=$2.4\% FWHM. Very similar resolutions are expected for different drift field configurations. In other words, further significant improvements in energy resolution for alpha particles with respect to the 2.8\% FWHM obtained here would require a better S1 detection efficiency and/or PMTs with better charge resolution.  

\begin{figure}[t!]
\begin{center}
\includegraphics[width=.49\textwidth]{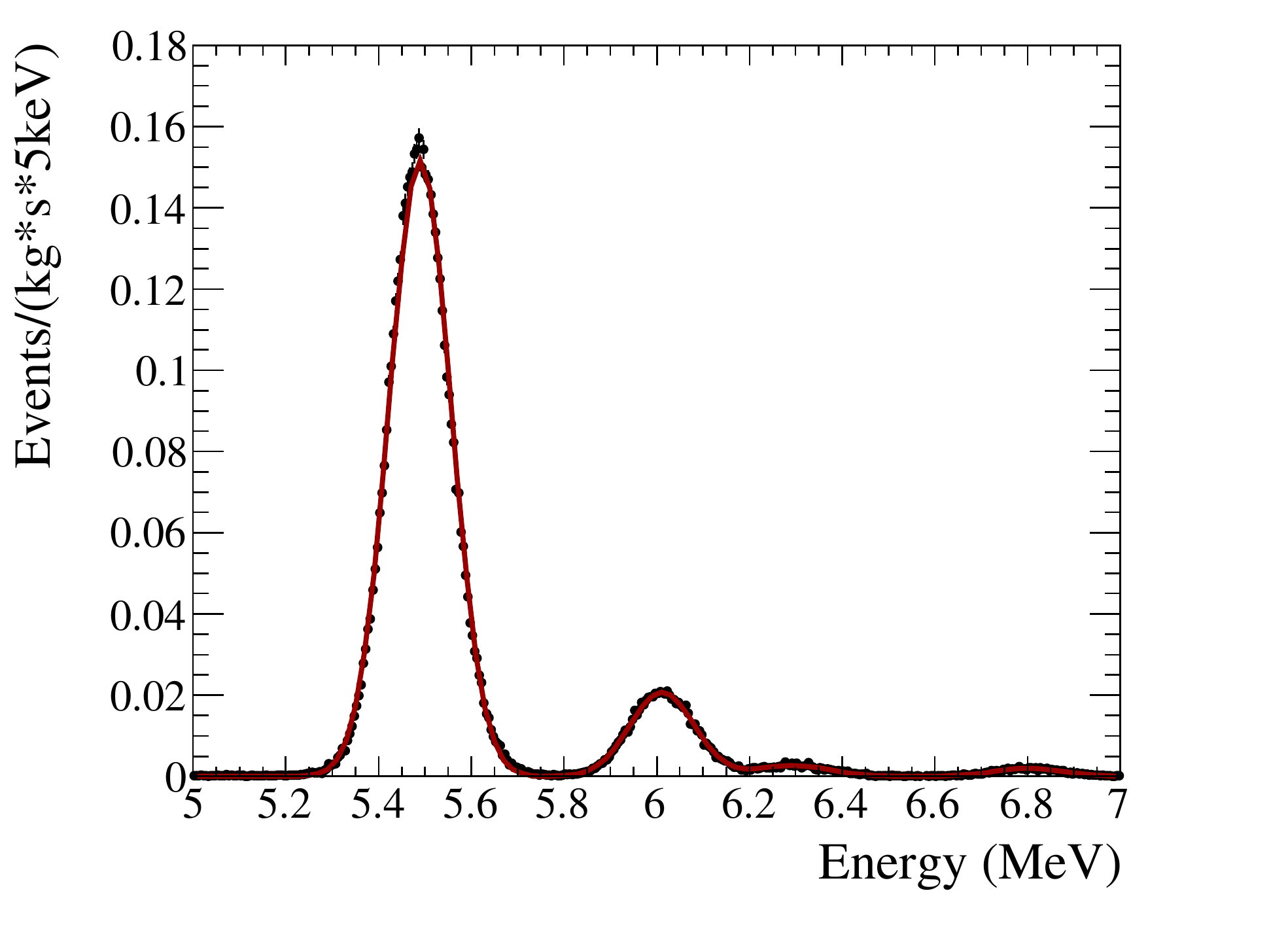} \hfill
\includegraphics[width=.49\textwidth]{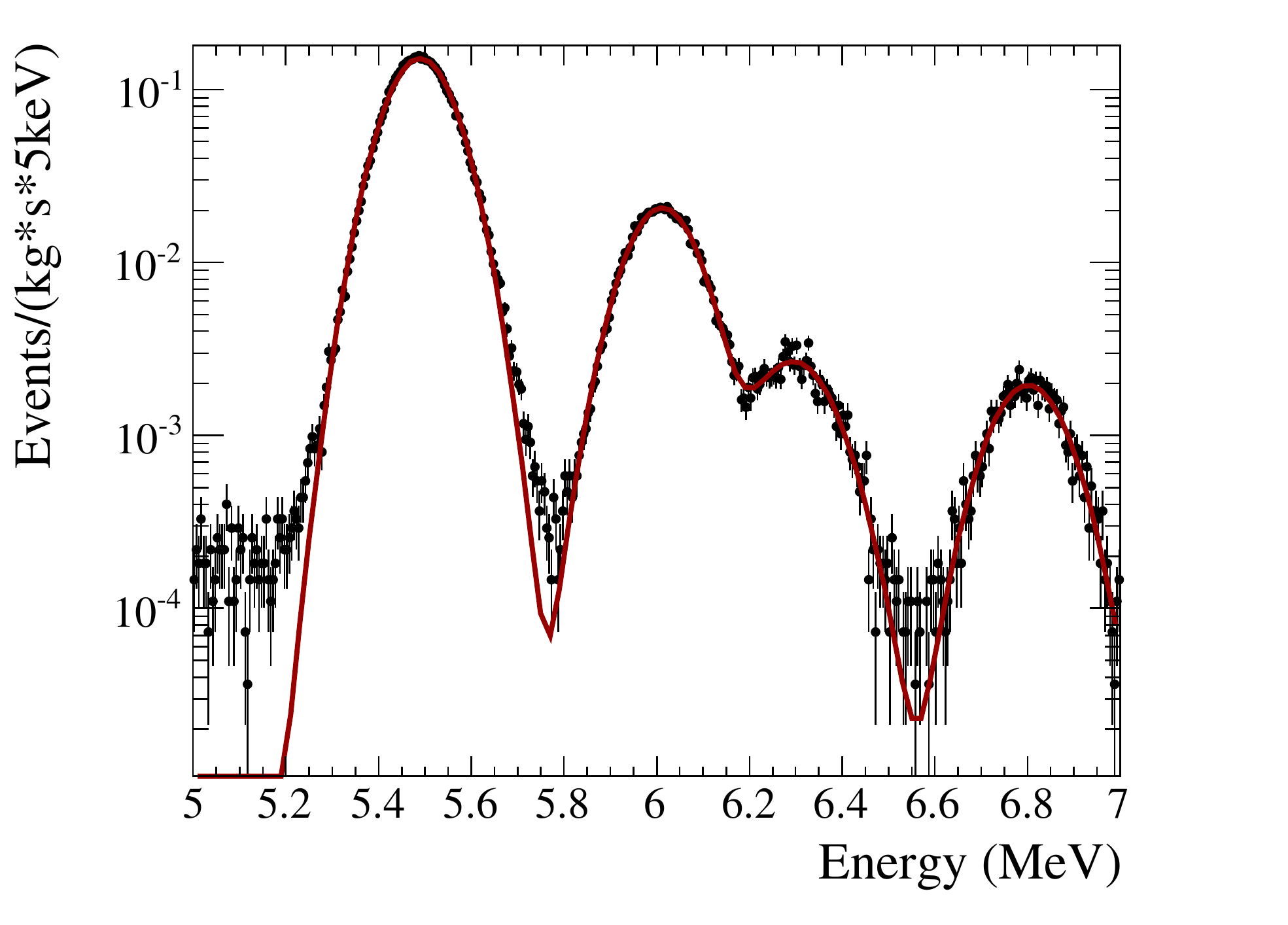}
\end{center}
\caption{Energy spectrum for alpha candidate events in NEXT-DEMO for the 1~kV/cm drift field run. The spectrum is shown with both linear and logarithmic vertical scales on the left and right panels, respectively. The energy is defined as in Eq.~\protect\ref{eq:energy}. The two lowest energy (and most intense) peaks correspond to \Rn{222}~(5.49~MeV) and \Po{218}~(6.00~MeV) from the alpha source, while the other two correspond to \Rn{220}~(6.29~MeV) and \Po{216}~(6.78~MeV) from natural impurities in the xenon gas. A fit to the data is also shown, as a sum of four gaussian functions.}\label{fig:energyspectrum}
\end{figure}

The event energy $E$ distribution for all alpha candidate events collected in the 1~kV/cm drift field run is shown in Fig.~\ref{fig:energyspectrum}. Four peaks are visible, corresponding (from left to right) to \Rn{222}~(5.49~MeV), \Po{218}~(6.00~MeV), \Rn{220}~(6.29~MeV) and \Po{216}~(6.78~MeV) alpha decays. The data are well described as the sum of four gaussian functions, also shown in the figure.

\subsection{Radon activity in the xenon gas} \label{subsec:ResultsActivity}

\begin{figure}[t!]
\begin{center}
\includegraphics[width=.49\textwidth]{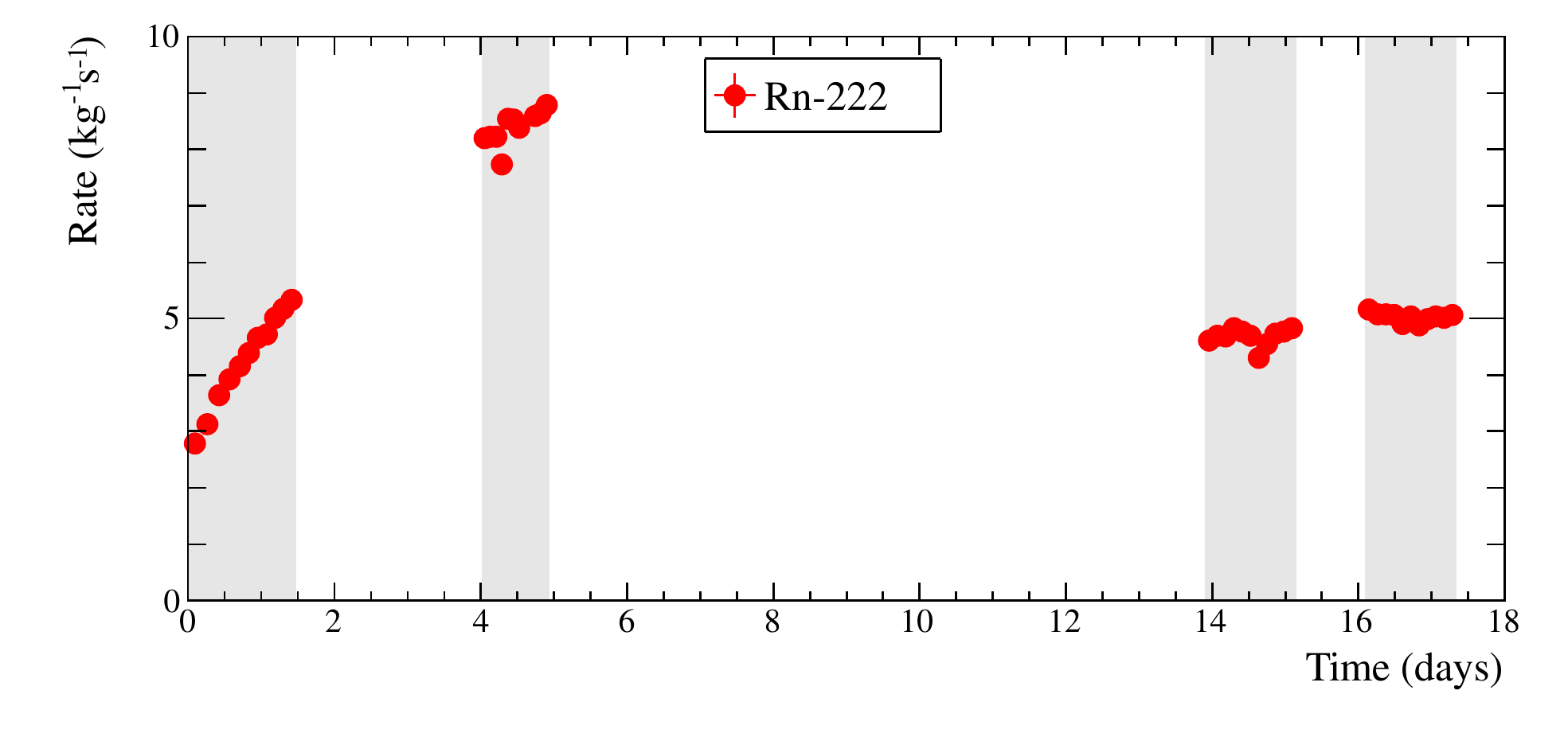} \hfill
\includegraphics[width=.49\textwidth]{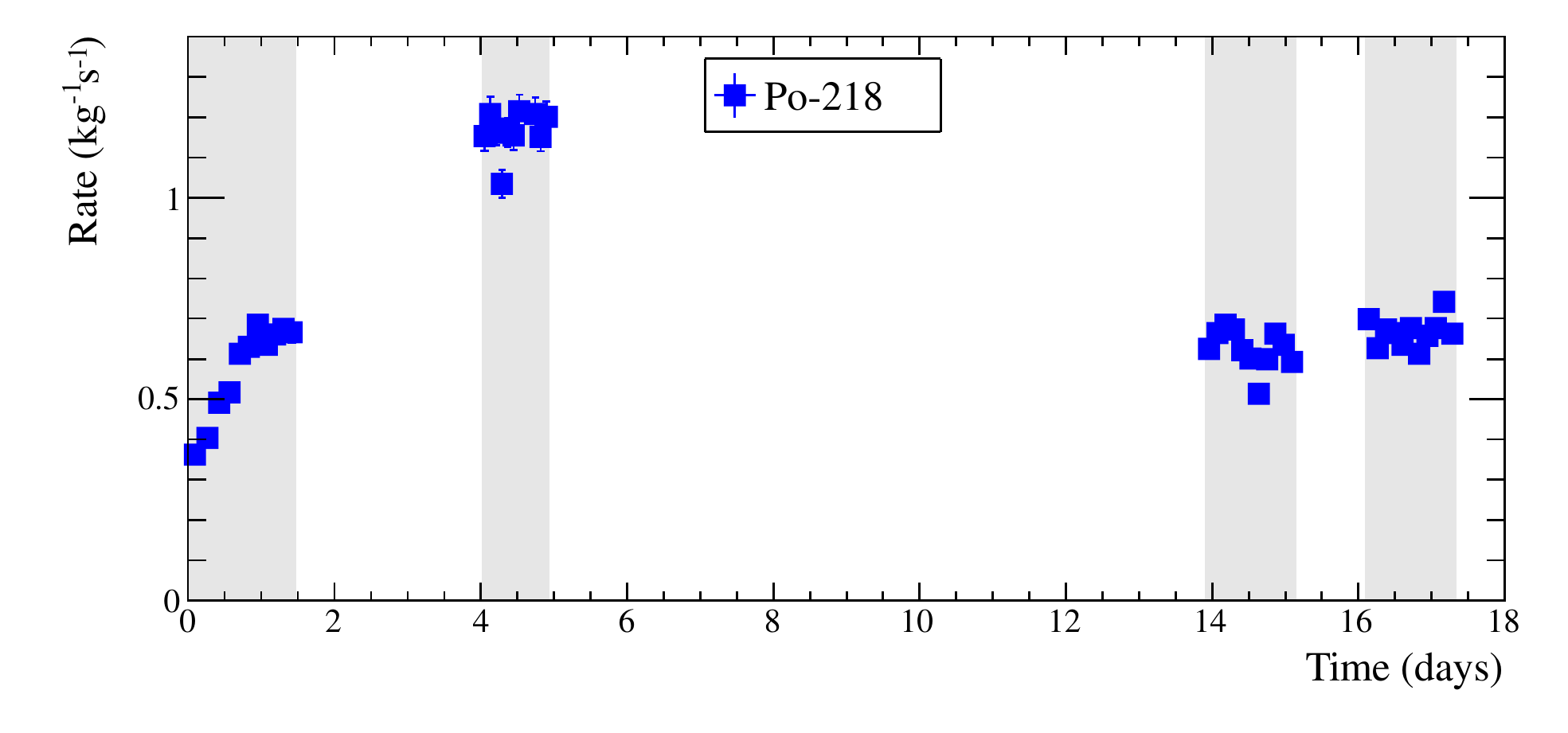}
\includegraphics[width=.49\textwidth]{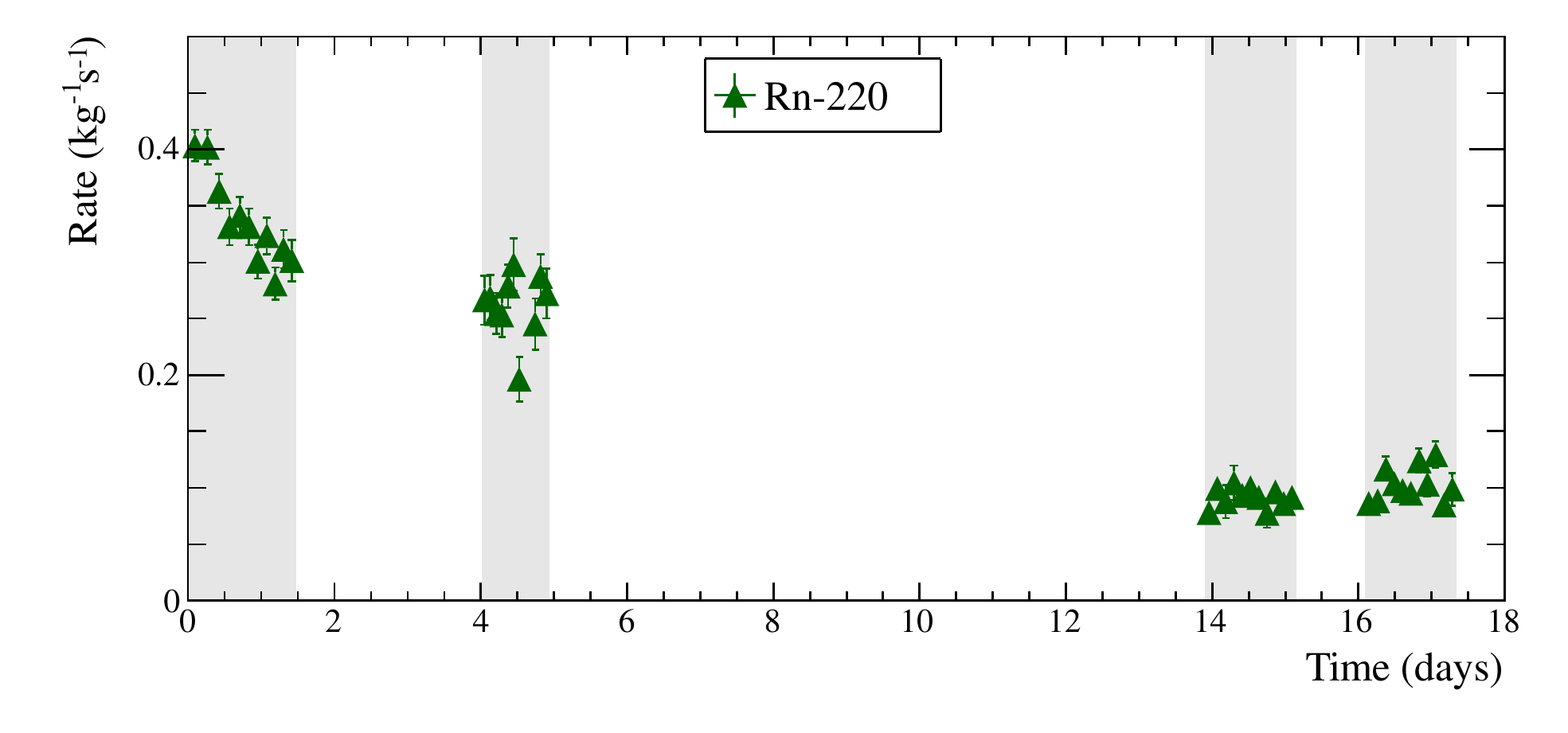} \hfill
\includegraphics[width=.49\textwidth]{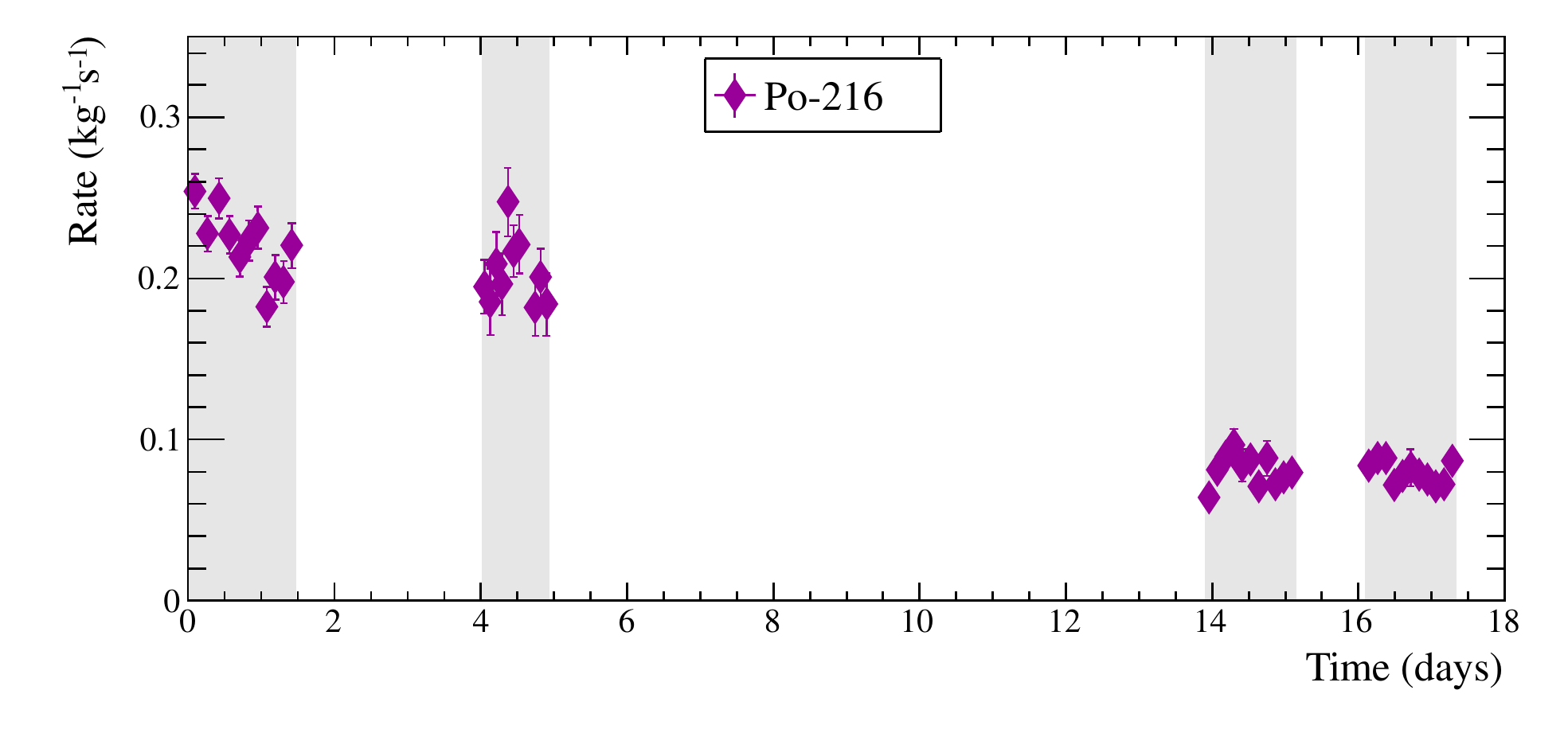}
\end{center}
\caption{Decay rate versus time for the four alpha-emitting isotopes \Rn{222}, \Po{218}, \Rn{220}, \Po{216}. The grey areas with data points correspond to the events used in this analysis, at 0.6, 0.3, 0.8 and 1~kV/cm drift field settings from early to late times, respectively. The time origin is set as the beginning of the 0.6~kV/cm run, approximately one day before the calibration source was inserted. The rate drop between the 0.6 and 0.8~kV/cm runs is due to changes in DAQ conditions.
}\label{fig:ratevstime}
\end{figure}

The gaussian fits to energy spectra as the one in Fig.~\ref{fig:energyspectrum} also allow us to estimate the alpha decay rate as a function of time for the four isotopes \Rn{222}, \Po{218}, \Rn{220}, and \Po{216}. The time evolution of the four decay rates is shown in Fig.~\ref{fig:ratevstime}, covering the entire data-taking period used in this work. The Radon source was introduced into the gas system one day before the time 0 shown in Fig.~\ref{fig:ratevstime}  (chosen as the start of the 0.6 kV/cm run) with gas circulating through the source continuously from that point. The rates clearly show that \Rn{222}\ and \Po{218}\ activities in the detector are produced by the source. Both rates increase by about a factor of 3 throughout our first two runs, at 0.6 and 0.3~kV/cm, taken across a time span of about 5 days. This is qualitatively consistent with the 3.8 day long half-life of \Rn{222}. 

On the other hand, we cannot draw any firm conclusion concerning the origin of \Rn{220}\ and \Po{216}\ in our detector. These isotopes could have been produced by \Rn{220}\ emanation from detector inner components, or from the room-temperature getter used during this run. An approximately constant \Rn{220}\ and \Po{216}\ decay rate would be expected in both cases. A third possibility is that \Rn{220}\ and \Po{216}\ are produced by the calibration source. These isotopes are not produced in the \Ra{226}\ decay chain, so any source-related \Rn{220}\ and \Po{216}\ activity would have to be produced by impurities in the source aluminum casing or in the source material itself. While source contamination appears unlikely, it cannot be excluded from our data, given that the short (56~s) long \Rn{220}\ half-life would imply an approximately constant \Rn{220}\ and \Po{216}\ decay rate also in this case. 

All four isotopes display a sudden drop in decay rate, by about a factor of two, between the 0.6 and 0.8~kV/cm runs. This drop was caused by instrumental effects, namely by a change in DAQ conditions. In order to operate in more stable DAQ conditions, the maximum allowed DAQ trigger rate was reduced from 10 to 5~Hz between the 0.6 and 0.8~kV/cm runs. The actual trigger rate in the last two runs was measured to be about 4~Hz, that is essentially at the DAQ saturation value. 

From Fig.~\ref{fig:ratevstime} and the above discussion, we conclude that the \Rn{220}\ specific activity originating from our experimental setup is at least 0.3~Bq/kg, and possibly even more if significant DAQ saturation were present during the 0.6 and 0.3~kV/cm runs. The measurement of the radon activity in the xenon gas is relevant for NEXT-100 \bbznu\ searches, since \Rn{220}\ and \Rn{222}\ decays ultimately produce \Tl{208}\ and \Bi{214}\ isotopes, respectively. Such isotopes either remain in the gas, or accumulate on the negatively charged electrostatic surfaces of the TPC before they are neutralized. The subsequent \Tl{208}\ and \Bi{214}\ decays produce gamma particles that are expected to be the dominant backgrounds to a \bbznu\ signal. We estimate that the specific activity for the radon in the gas measured in the non-radiopure NEXT-DEMO detector, if simply extrapolated to the NEXT-100 geometry, would yield a \bbznu\ background that is about 3 orders of magnitude higher than any other background source expected in NEXT-100 \cite{Gomez-Cadenas:2013lta}. While this is intolerably high, we expect to be able to reduce this contribution to a negligible level in a radiopure detector operated underground such as NEXT-100. For example, the EXO-200 detector has measured a radon (in this case, \Rn{222}) specific activity in liquid xenon of 3.65$\pm$0.37~$\mu$Bq/kg \cite{Albert:2013gpz}, yielding a negligible background contribution for EXO-200 \bbznu\ searches. In any case, the analysis techniques described in this work could be applied to provide an accurate monitoring of the amount of radon that is present in the NEXT-100 gas system.

\subsection{Field dependence of ionization and scintillation yields} \label{subsec:ResultsFieldDependence}

\begin{figure}[t!]
\begin{center}
\includegraphics[width=.65\textwidth]{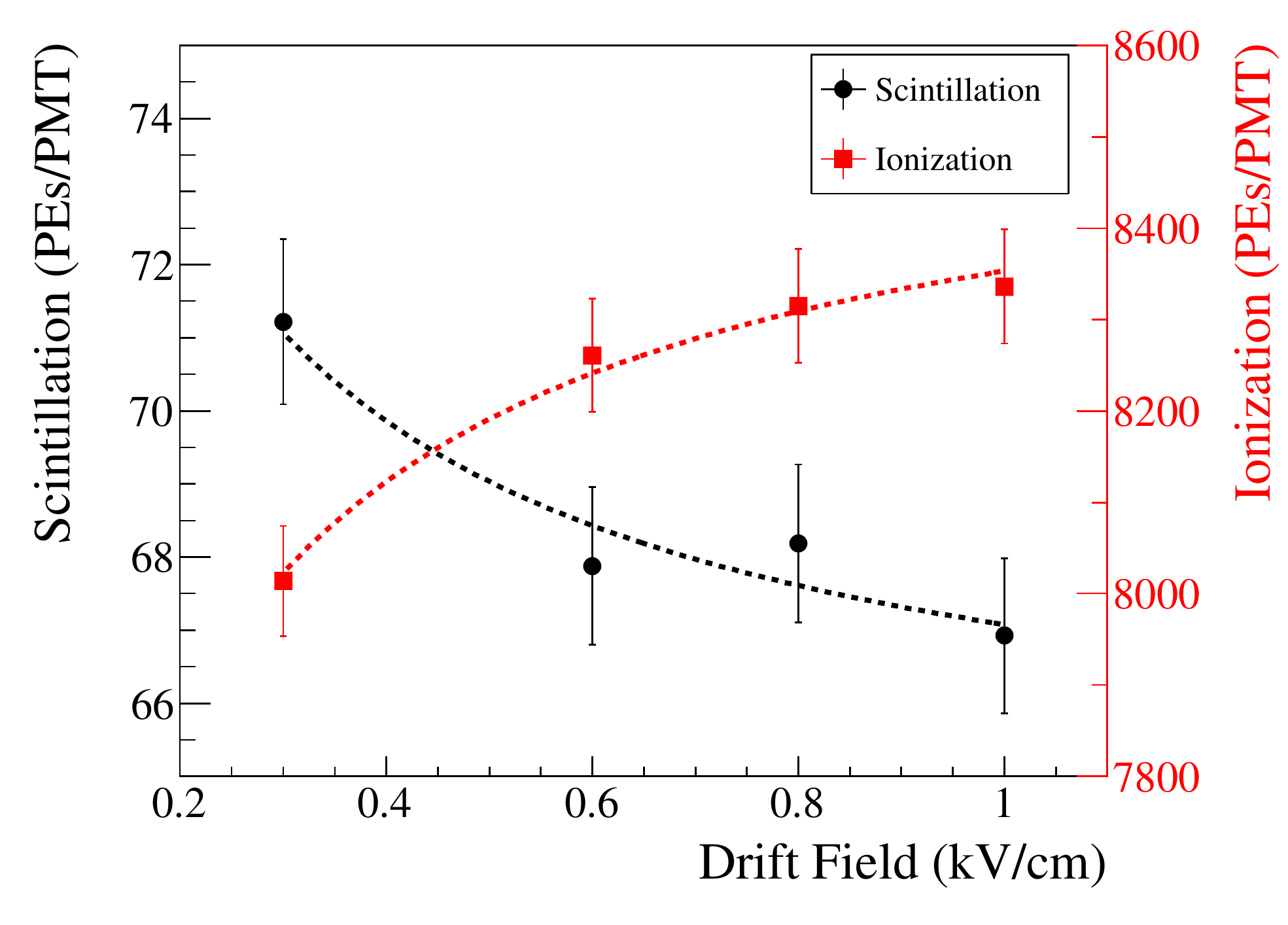} 
\end{center}
\caption{Average primary scintillation (black circles) and ionization (red squares) signals for \Rn{222}\ alpha candidate events only, as a function of drift field. The dashed lines indicate the best-fit curves for the two-component recombination model given by Eqs.~\protect\ref{eq:n2yieldvsfield} and \protect\ref{eq:n1yieldvsfield}.}\label{fig:means1s2vsfield}
\end{figure}

Electron-ion recombination also produces a characteristic drift field dependence for the average ionization and primary scintillation yields. In this work, we study such dependence between 0.3 and 1~kV/cm fields, for \Rn{222} (5.49~MeV) alpha candidate events only. As the drift field is increased, electron-ion recombination is reduced on average, causing \None\ to decrease and \Ntwo\ to increase. This expectation is confirmed by our measurements in Fig.~\ref{fig:means1s2vsfield}. The S1 error bars in Fig.~\ref{fig:means1s2vsfield} are given by the uncertainty in extrapolating the \None\ yields to $z=0$, while the S2 uncertainties are dominated by \ELGain\ variations caused by the 0.7\% gas pressure variations during data-taking.

The dashed curves in Fig.~\ref{fig:means1s2vsfield} indicate the expectation that fits our data best from the same two-component recombination model used in \cite{bolotnikov1999, Alvarez:2012hu}. The average ionization yield is parametrized via:
\begin{equation}
\langle\Ntwo\rangle \equiv Q(E_{\rm drift}) = Q_0[0.8+0.2(1+K_2/E_{\rm drift})^{-1}]
\label{eq:n2yieldvsfield}
\end{equation}
\noindent where $E_{\rm drift}$ is the drift field, and $K_2$ and $Q_o$ are fit parameters: the parameter that controls the drift field dependence of (geminate) recombination and the total ionization in absence of recombination (at infinite drift field), respectively. As in \cite{Alvarez:2012hu}, we fix the fraction of total charge that may undergo geminate recombination to $Q_2/Q_0 = 0.2$, so that $Q_1/Q_0 = 0.8$ is the fixed fraction that may undergo columnar recombination (but only for drift fields $\lesssim 0.1$ kV/cm, see \cite{bolotnikov1999}). The average primary scintillation yield is fitted with:
\begin{equation}
\langle\None\rangle \equiv L(E_{\rm drift}) = L_{\rm ex}[1+1.6(1-Q(E_{\rm drift})/Q_0)]
\label{eq:n1yieldvsfield}
\end{equation}
\noindent where $Q(E_{\rm drift})$ is given by Eq.~\ref{eq:n2yieldvsfield}, and $L_{\rm ex}$ is a third fit parameter, namely the scintillation light yield in absence of recombination. As in \cite{Alvarez:2012hu}, we make a further simplifying assumption: we take the ratio of $L(E_{\rm drift})-L_{\rm ex}$ to $Q_0-Q(E_{\rm drift})$ to be approximately equal to $1.6\cdot L_{\rm ex}/Q_0$, as motivated by \cite{Saito:2003dz}. 

Overall, our recombination model for describing both the charge and light yields (Eqs.~\ref{eq:n2yieldvsfield} and \ref{eq:n1yieldvsfield}) include three fit parameters: $L_{\rm ex}$, $Q_0$ and $K_2$. The fit to our data yields $L_{\rm ex}=(64.6\pm 1.1)$ PEs/PMT, $Q_0=(8559\pm 98)$ PEs/PMT, $K_2=(0.137\pm 0.048)$ kV/cm, with a goodness-of-fit of $\chi^2$/dof of 0.8/5. The data are therefore consistent with the model, at least within our relatively large uncertainties. We estimate that a fraction $r=1-\langle\Ntwo\rangle/Q_0=0.026\pm 0.013$ of ionization charge undergoes recombination at 1~kV/cm drift field. This fraction increases to $r=0.064\pm 0.013$ at 0.3~kV/cm field. 

\subsection{Energy partitioning between ionization and scintillation} \label{subsec:ResultsWex}

As discussed in Sec.~\ref{sec:Introduction}, the ratio \Nex/\Ni\ of excited to ionized atoms produced in xenon gas by ionizing radiation (in our case, by alpha particles), can be estimated from the optical gain $\ELGain=(75\pm 5)$, the primary scintillation and ionization yields \None\ and \Ntwo, and from the electron-ion recombination probability $r$, see Eq.~\ref{eq:nexniratio2}. We compute the ratio for the 1~kV/cm drift field data set, the least affected by recombination corrections. In this case, by taking $\None = (66.93\pm 1.06)$~PEs/PMT, $\Ntwo = (8336\pm 63)$~PEs/PMT and $r = (0.026\pm 0.013)$, we obtain $\Nex/\Ni = (0.561\pm 0.045)$. Our measured value is therefore consistent with previous \Nex/\Ni\ results in high-pressure xenon gas using alpha particles, see Sec.~\ref{sec:Introduction} and references \cite{Saito:2003dz,mimura} therein.

Furthermore, by considering that the average energy \Wi\ to produce one electron-ion pair has been accurately measured to be about 22~eV, our \Nex/\Ni\ ratio can be translated into a measurement of \Wex, the average energy to produce one excited atom in xenon gas:
\begin{equation}
\Wex = \frac{\Wi}{(\Nex/\Ni)}= (39.2\pm 3.2)~{\rm eV}
\label{eq:Wex}
\end{equation}
Our measured \Wex\ value can be compared with previous measurements of the same quantity, see Tab.~\ref{tab:Wex}. Again, our value is compatible with the results obtained with alpha particles in \cite{Saito:2003dz,mimura}. On the other hand, our \Wex\ measurement is lower than the values obtained in \cite{docarmo}, \cite{Fernandes:2010gg}, \cite{Parsons:1990hv} and \cite{Renner:2014mha}, which use X-ray and gamma ray sources.

\section{Conclusions} \label{sec:Conclusions}

In this work we have studied the mechanisms of electron-ion recombination and of energy loss partitioning between ionization and excitation, in a time projection chamber filled with xenon gas at 10~bar pressure. We have used the NEXT-DEMO prototype of the NEXT-100 neutrinoless double beta decay detector, recently upgraded with a silicon photomultiplier tracking array \cite{Alvarez:2013gxa}. The detector was exposed to a \Ra{226}\ calibration source, and alpha particles from \Rn{222}~(5.49~MeV) and \Po{218}~(6.00~MeV) decay have been selected. Both the primary (prompt, or S1) scintillation and the ionization yields produced by alpha particles have been studied. The ionization signal is converted into secondary (delayed, or S2) scintillation light via an electroluminescent amplification stage, and detected via the same photo-detectors that record S1 light. Alpha decays throughout the NEXT-DEMO active volume have been used, after correcting for detector-dependent effects affecting the charge and light yields as a function of alpha particle spatial position within the gas. 

Alpha particles are sufficiently ionizing to produce sizable electron-ion recombination effects in xenon even in its gaseous form, and for the 0.3--1~kV/cm drift field intensities studied in this work. As previously observed in liquid xenon, we have measured event-by-event correlation fluctuations between ionization and scintillation due to recombination, since every recombined electron produces additional scintillation, as schematically shown in Fig.~\ref{fig:detctionprocess}. We have verified that this charge-light anti-correlation is strongest at low drift fields, where recombination effects are more important, as shown in Fig.~\ref{fig:s2vss1}. In particular, we have measured S1-S2 correlation coefficients between -0.80 (at 0.3~kV/cm) and -0.56 (at 1~kV/cm) for \Rn{222}\ alpha particles. This anti-correlation can be exploited to define an improved energy estimator that combines both charge and light information. In an electroluminescent TPC such as NEXT-DEMO, the linear combination of S1 and S2 signals providing the best energy resolution is expected to be the one where the S2 signal is weighted by the inverse of the effective electroluminescence (or optical) gain of the detector. Using \Rn{222}\ events only, we determine the best energy energy resolution to be 2.8\% FWHM at an alpha particle kinetic energy of 5.49~MeV, and the effective optical gain to be $(75\pm 5)$ S2 photons per ionization electron, see Figs.~\ref{fig:s1s2profile} and \ref{fig:energyresolution}. We have also studied how electron-ion recombination affects the average scintillation and ionization yields as a function of drift field intensity over the 0.3--1~kV/cm drift field range. The observed increase in charge, and corresponding decrease in light, as a function of increasing drift field is well described by the same two-component recombination model used in \cite{bolotnikov1999}, at least within our relatively large uncertainties. In this model, both geminate \cite{Onsager:1938zz} and columnar \cite{jaffe} recombination effects are considered.

The NEXT-DEMO prototype, instrumented with photo-detectors that simultaneously measure the ionization-induced and the excitation-induced signals, is ideally suited to characterize also the different pathways of energy loss in xenon gas. In particular, we measure the ratio of excited to ionized atoms in xenon gas produced by alpha particles to be $0.561\pm 0.045$, consistent with previous measurements in high-pressure xenon gas and using alpha particles \cite{Saito:2003dz,mimura}. This ratio measurement can be combined with the accurately known ionization yield \Wi, about 22~eV to produce an electron-ion pair, to obtain a measurement of \Wex, the average energy required to excite one xenon atom. The quantity \Wex\ is subject to considerably more uncertainty compared to \Wi, (see Tab.~\ref{tab:Wex} for previous measurements in xenon gas). We measure \Wex\ to be $(39.2\pm 3.2)$~eV, again consistent with earlier results with alpha calibration sources \cite{Saito:2003dz,mimura}, but lower than what obtained with X-ray and gamma ray projectiles \cite{docarmo,Fernandes:2010gg,Parsons:1990hv,Renner:2014mha}.

Compared to our previous studies in \cite{Alvarez:2012hu}, the upgraded NEXT-DEMO detector and the improved analysis techniques also allowed us to select alpha particles that are unrelated to the dominant \Rn{226}\ decay chain and that are produced by radioactive impurities in our experimental setup, namely from \Rn{220}~(6.29~MeV) and \Po{216}~(6.78~MeV) isotopes. We have measured the specific activity of \Rn{220}\ in the xenon gas to be at least 0.3~Bq/kg in the non-radiopure NEXT-DEMO detector. The same alpha spectroscopy techniques developed in this work could be applied to the radiopure NEXT-100 detector to provide an accurate monitoring of the amount of radon that is present inside the xenon gas, thus measuring a potential background source for neutrinoless double beta decay searches.


\acknowledgments
This work was supported by the following agencies and institutions: the European Research Council under the Advanced Grant 339787-NEXT; the Ministerio de Econom\'ia y Competitividad of Spain under grants CONSOLIDER-Ingenio 2010 CSD2008-0037 (CUP), FPA2009-13697-C04 and FIS2012-37947-C04; the Director, Office of Science, Office of Basic Energy Sciences, of the US Department of Energy under contract no.\ DE-AC02-05CH11231; and the Portuguese FCT and FEDER through the program COMPETE, project PTDC/FIS/103860/2008.

\end{document}